%% file: Vazquez-Bare_arxiv.tex
\documentclass[12pt]{article}
\usepackage[onehalfspacing]{setspace}
\usepackage[margin=0.8in]{geometry}
\usepackage{amsmath,amsfonts,amssymb,mathrsfs}
\usepackage{bbm}
\usepackage{graphicx,subcaption}
\usepackage{hyperref}
\usepackage[round]{natbib}
\usepackage{ntheorem}
\usepackage[title]{appendix}
\usepackage{enumerate}
\usepackage{xr}
\usepackage{pdflscape}
\usepackage{afterpage}

\hypersetup{
    colorlinks=true, 
    linktoc=all,        
    linkcolor=blue,  
    citecolor=blue,
    urlcolor=blue
}

\renewcommand{\P}{\mathbb{P}}
\newcommand{\E}{\mathbb{E}}
\newcommand{\V}{\mathbb{V}}
\newcommand{\cov}{\mathbb{C}\text{ov}}
\newcommand{\1}{\mathbbm{1}}

\newtheorem{example}{Example}
\newtheorem{remark}{Remark}
\newtheorem{definition}{Definition}
\newtheorem{theorem}{Theorem}

\newtheorem{lemma}{Lemma}

\newtheorem{assumption}{Assumption}

\newtheorem{corollary}{Corollary}

\newcommand\indep{\protect\mathpalette{\protect\independenT}{\perp}}
\def\independenT#1#2{\mathrel{\rlap{$#1#2$}\mkern2mu{#1#2}}}

\title{Identification and Estimation of Spillover Effects \\ in Randomized Experiments\footnote{I am deeply grateful to Matias Cattaneo for continued advice and support. I am indebted to Lutz Kilian, Mel Stephens and Roc{\'i}o Titiunik for thoughtful feedback and discussions. I thank Cl{\'e}ment de Chaisemartin, Catalina Franco, Amelia Hawkins, Nicol{\'a}s Idrobo,  Xinwei Ma, Nicolas Morales, Kenichi Nagasawa, Olga Namen and Doug Steigerwald for valuable discussions and suggestions, and seminar participants at UChicago, UCLA, UC San Diego, University of Michigan, UC Santa Barbara, UChicago Harris School of Public Policy, Cornell University, UChicago Booth School of Business, UT Austin, Stanford University and UC Berkeley for helpful comments. I also thank the editor, Elie Tamer, the associate editor and three anonymous referees for their detailed comments and suggestions that greatly improved the paper.}}

\author{Gonzalo Vazquez-Bare\thanks{Department of Economics, University of California, Santa Barbara. \href{gvazquez@econ.ucsb.edu}{gvazquez@econ.ucsb.edu}.}}

\begin{document}
\maketitle

\vspace{.5cm}

\abstract{I study identification, estimation and inference for spillover effects in experiments where units’ outcomes may depend on the treatment assignments of other units within a group. I show that the commonly-used reduced-form linear-in-means regression identifies a weighted sum of spillover effects with some negative weights, and that the difference in means between treated and controls identifies a combination of direct and spillover effects entering with different signs. I propose nonparametric estimators for average direct and spillover effects that overcome these issues and are consistent and asymptotically normal under a precise relationship between the number of parameters of interest, the total sample size and the treatment assignment mechanism. These findings are illustrated using data from a conditional cash transfer program and with simulations. The empirical results reveal the potential pitfalls of failing to flexibly account for spillover effects in policy evaluation: the estimated difference in means and the reduced-form linear-in-means coefficients are all close to zero and statistically insignificant, whereas the nonparametric estimators I propose reveal large, nonlinear and significant spillover effects.
}

\vspace{1cm}


\vspace{1cm}


\setcounter{page}{0}\thispagestyle{empty}

\newpage


\section{Introduction}

Spillover effects, which occur when an agent's actions or behaviors indirectly affect other agents' outcomes through peer effects, social interactions or externalities, are ubiquitous in economics and social sciences. A thorough account of spillover effects is crucial to assess the causal impact of policies and programs \citep{Abadie-Cattaneo_2018_Annurev,Athey-Imbens_2017_handbook}. However, the literature is still evolving in this area, and most of the available methods for analyzing treatment effects either assume no spillovers or allow for them in restrictive ways, often without a precise definition of the parameters of interest or the conditions required to recover them. 

This paper studies identification, estimation and inference for average direct and spillover effects in randomized controlled trials, and offers three main contributions. First, I provide conditions for nonparametric identification of causal parameters when the true spillovers structure is possibly unknown. Under the assumption that interference occurs within non-overlapping peer groups, I define a rich set of direct and spillover treatment effects based on a function, the \textit{treatment rule}, that maps peers' treatment assignments and outcomes. Lemma \ref{lemma:identif} links average potential outcomes to averages of observed variables when the posited treatment rule is possibly misspecified.

The second main contribution is to characterize the difference in means between treated and controls, and the coefficients from a reduced-form linear-in-means (RF-LIM) regression, two of the most commonly analyzed estimands when analyzing RCTs and spillover effects in general. Theorem \ref{thm:dm} shows that, in the presence of spillovers, the difference in means between treated and controls combines the direct effect of the treatment and the difference in spillover effects for treated and untreated units, and thus the sign of the difference in means is undetermined even when the signs of all direct and spillover effects are known. On the other hand, Theorem \ref{thm:lim} shows that a RF-LIM regression recovers a linear combination of spillover effects for different numbers of treated peers where the weights sum to zero, and hence some weights are necessarily negative. As a result, the coefficients from a RF-LIM regression can be zero even when all the spillover effects are non-zero. I then provide sufficient conditions under which the difference in means and the RF-LIM coefficients have a causal interpretation, that is, when they can be written as proper weighted averages of direct and/or spillover effects. I also propose a simple regression-based pooling strategy that is robust to nonlinearities and heterogeneity in spillover effects. 

The third main contribution is to analyze nonparametric estimation and inference for spillover effects. In the presence of spillovers, the number of treatment effects to estimate can be large, and the probability of observing units under different treatment assignments can be small. Section \ref{sec:estimation} provides general conditions for uniform consistency and asymptotic normality of the estimators of interest in a double-array asymptotic framework where both the number of groups and the number of parameters are allowed to grow with the sample size. This approach highlights the role that the number of parameters and the assignment mechanism play on the asymptotic properties of nonparametric estimators. More precisely, consistency and asymptotic normality are shown under two main conditions that are formalized in the paper: (i) the number of treatment effects should not be ``too large'' with respect to the sample size, and (ii) the probability of each treatment assignment should not be ``too small''. These two requirements are directly linked to modeling assumptions on the potential outcomes, the choice of the set of parameters of interest and the treatment assignment mechanism. As an alternative approach to inference, the wild bootstrap is shown to be consistent, and simulation evidence suggests that it can yield better performance compared to the normal approximation in some settings.

The results in this paper are illustrated in a simulation study and using data from a randomized conditional cash transfer. The empirical results clearly highlight the pitfalls of failing to flexibly account for spillovers in policy evaluation: the estimated difference in means and RF-LIM coefficients are all close to zero and statistically insignificant, whereas the nonparametric estimators I propose reveal large, nonlinear and significant spillover effects. 

This paper is related to a longstanding literature on peer effects and social interactions. A large strand of this literature has focused on identification of social interaction effects in parametric models. The most commonly analyzed specification is the linear-in-means (LIM) model, where a unit's outcome is modeled as a linear function of own characteristics, peers' average characteristics and peers' average outcomes \citep[see e.g.][for recent reviews]{Blume-etal_2015_JPE,Kline-Tamer_2019,Bramoulle-Djebbari-Fortin_2020_AnnuRev}. Since \citet{Manski_1993_Restud}'s critique of LIM models, several strategies have been put forward to identify structural parameters \citep[see e.g.][]{Lee_2007_JoE,Bramoulle-Djebbari-Fortin_2009_JoE,Davezies-etal_2009_EJ,DeGiorgi-Pellizzari-Redaelli_2010_AEJ}. All these strategies rely on linearity of spillover effects. In this paper, I consider an alternative approach that focuses on reduced-form casual parameters from a potential-outcomes perspective. Within this setup, I show that (reduced-form) response functions can be identified and estimated nonparametrically. While I do not consider identification of structural parameters in this paper, reduced-form parameters are inherently relevant, as they represent the causal effect of changing the peers' covariate values, which is generally more easily manipulable for a policy maker than peers' outcomes \citep{Goldsmith-Pinkham-Imbens_2013_JBES,Manski_2013_JBES}. This is particularly true in my setup, where the covariate of interest is a treatment that is assigned by the policy maker. Furthermore, identification of reduced-form parameters can be thought of as a necessary ingredient for identifying structural models.

On the opposite end of the spectrum, \citet{Manski_2013_EJ} and \citet{Lazzati_2015_QE} study nonparametric partial identification of response functions under different restrictions on the structural model, the response functions and the structure of social interactions. My paper complements this important strand of the literature by considering a specific network structure in which spillovers are limited to non-overlapping groups, where the within-group spillovers structure is left unrestricted. This specific structure of social interactions allows me to obtain point (as opposed to partial) identification of causal effects without knowing the true mapping between treatment assignments an potential outcomes in a setting with wide empirical applicability. Furthermore, by focusing on this network structure I can analyze the effect of misspecifying the within-group spillovers structure, as discussed in Section \ref{sec:identification}. Finally, I also complement this literature by providing a formal treatment of estimation and inference and showing validity of the wild bootstrap.

Another body of research has analyzed causal inference under interference from a design-based perspective in which potential outcomes are fixed and all randomness is due to the (known) treatment assignment mechanism \citep[see][for reviews]{Tchetgen-VdW_2012_SMMR,Ogburn-VanderWeele_2014_SS,Halloran-Hudgens_2016_CER}. Generally, this literature focuses on two aggregate measures of treatment effects, the average direct effect and the average indirect effect, that average over peers' assignments under specific treatment assignment mechanisms such as completely randomized designs \citep{Sobel_2006_JASA} or two-stage randomization \citep[][and subsequent studies]{Hudgens-Halloran_2008_JASA}. In a related paper, \citet{Athey-Eckles-Imbens_2017_JASA} derive a procedure to calculate finite-sample randomization-based p-values to test for the presence of spillover effects. My paper complements this literature in several ways. First, I focus on identifying and estimating the entire vector of spillover effects determined by the treatment rule, which can be seen either as the main object of interest, or as an ingredient to construct the aggregate summary measures of spillovers considered in the literature (see also Remark \ref{rmk:pooled}). Second, my identification results allow for general treatment assignment mechanisms. Third, I consider large-sample inference from a super-population perspective that allows me to disentangle the roles of the number of groups, the number parameters of interest and the treatment assignment mechanism in the performance of estimators and inferential procedures.

Finally, \citet{Moffit_2001}, \citet{Duflo-Saez_2003_QJE}, \citet{,Hirano-Hahn_2010_EL} and more recently \citet{Baird-etal_2018_Restat} analyze the design of partial population experiments, where spillovers are estimated by exposing experimental units to different proportions of treated peers (or ``saturations''). My paper complements this strand of the literature by analyzing identification under general experimental designs and by providing a formal treatment of the effect of the experimental design on inference, which formalizes the advantages of partial population designs. This fact is illustrated in Section \ref{sec:simulation} and discussed in more detail in Section \ref{sec:design} of the supplemental appendix. 

The remainder of the paper is organized as follows. Section \ref{sec:setup} describes the setup and defines the parameters of interest. Section \ref{sec:identification} provides the main identification results. Section \ref{sec:estimation} analyzes estimation and inference. Section \ref{sec:simulation} provides a simulation study, and Section \ref{sec:empapp} contains the empirical application. Section \ref{sec:conclusion} concludes. The proofs, together with additional results and discussions, are provided in the supplemental appendix. 


\section{Setup}\label{sec:setup}

As a motivating example, consider a program in which parents in low-income households receive a cash transfer from the government provided their children are enrolled in school and reach a required level of attendance. Suppose that this conditional cash transfer program is evaluated using a randomized pilot in which children are randomly selected to participate. There are several reasons to expect within-household spillovers from this program. On the one hand, the cash transfer may alleviate financial constraints that were preventing the parents from sending their children to school on a regular basis. The program could also help raise awareness on the importance of school attendance. In both these cases, untreated children may indirectly benefit from the program when they have a treated sibling. On the other hand, the program could create incentives for the parents to reallocate resources towards their treated children and away from their untreated siblings, decreasing school attendance for the latter. In all cases, ignoring spillover effects can severely underestimate the costs or the benefits of this policy. 

Moreover, these alternative scenarios have drastically different implications on how to assign the program when scaling it up. In the first two situations, treating one child per household can be a cost-effective way to assign the treatment, whereas in the second case, treating all the children in a household can be more beneficial. An accurate assessment of spillovers is therefore crucial for the analysis and design of public policies.


\subsection{Notation and parameters of interest}

Consider a random sample of groups indexed by $g=1,\ldots,G$, each with $n_g+1$ units, so that each unit $i$ in group $g$ has $n_g$ neighbors or peers and $1\le n_g<\infty$. I assume group membership is observable. Units in each group are assigned a binary treatment, and a unit's potential outcomes, defined below, can depend on the assignment of all other units in the same group. Using the terminology of \citet{Ogburn-VanderWeele_2014_SS}, this phenomenon is known as \textit{direct interference}. Interference is assumed to occur between units in the same group, but not between units in different groups.

The individual treatment assignment of unit $i$ in group $g$ is denoted by $D_{ig}$, taking values $d\in\{0,1\}$, and the vector of treatment assignments in each group is given by $\mathbf{D}_g=(D_{1g},\ldots,D_{n_g+1,g})$. For each unit $i$, $D_{jig}$ is the treatment indicator corresponding to unit $i$'s $j$-th neighbor, collected in the vector $\mathbf{D}_{(i)g}=(D_{1ig},D_{2ig},\ldots,D_{n_gig})$. This vector takes values $\mathbf{d}_g=(d_1,d_2,\ldots,d_{n_g})\in \mathcal{D}\subseteq \{0,1\}^{n_g}$. 

A key element in this setup will be a function $h_0(\cdot)$ that summarizes how the vector $\mathbf{d}_g$ enters the potential outcome. More precisely, define a function or \textit{treatment rule}:
\[h_0:\mathcal{D}\to\mathcal{H}_0\]
that maps $\mathbf{d}_g$ into some value $h_0(\mathbf{d}_g)$ of the same or smaller dimension, so that $\mathrm{dim}(\mathcal{H}_0)\le \mathrm{dim}(\mathcal{D})$. Following \cite{Manski_2013_EJ}'s terminology, for $h_0(\mathbf{d}_g)=\mathbf{h}_0$, I will refer to the tuple $(d,\mathbf{h}_0)$ as the \textit{effective treatment assignment}, an element in the set $\{0,1\}\times\mathcal{H}_0$. The potential outcome for unit $i$ in group $g$ is denoted by the random variable $Y_{ig}(d,\mathbf{h}_0)$ where $\mathbf{h}_0=h_0(\mathbf{d}_g)\in\mathcal{H}^0$.

\begin{example}[SUTVA]
\upshape If $h_0(\cdot)$ is a constant function, the vector of peers' assignments is ignored and the set of effective treatment assignments becomes $\{0,1\}$, so the potential outcomes do not depend on peers' assignments. In this case the only effective treatment assignments are $D_{ig}=1$ and $D_{ig}=0$ (treated and control). This assumption is often known as the stable unit treatment value assumption or SUTVA \citep{Imbens-Rubin_2015_book}. $\square$
\end{example}

\begin{example}[Exchangeability]\label{ex:exchange}
\upshape When potential outcomes depend on how many peers, but not which ones, are treated,  peers are said to be \textit{exchangeable}. Exchangeability can be modeled by setting $h_0(\mathbf{d}_g)=\mathbf{1}_g'\mathbf{d}_g$ so $h_0(\cdot)$ summarizes $\mathbf{d}_g$ through the sum of its elements. The set of effective treatment assignments in this case is given by $\{(d,s):d=0,1,s=0,1,\ldots,n_g\}$. Exchangeability may be a natural starting point when there is no clear way (or not enough information) to assign identities to peers. $\square$
\end{example}

\begin{example}[Stratified exchangeability]
\upshape Exchangeability may also be imposed by subgroups. For instance, the vector of assignments may be summarized by the number of male and female treated peers separately. In settings where units are geographically located, peers are commonly assumed to be exchangeable within groups defined by distance such as within one block, between one and two blocks, etc, or by different distance radiuses (e.g. within 100 meters, between 100 and 200 meters and so on). $\square$
\end{example}

\begin{example}[Reference groups]
\upshape When each unit interacts only with a strict subset of her peers, we can define for example $h_0(\cdot):\{0,1\}^{n_g}\to \{0,1\}^{k_g}$ where $k_g<n_g$. For instance, under the assumption that each unit interacts with her two closest neighbors, $h_0(\mathbf{d}_g)=(d_1,d_2)$ so that $k_g=2$. The subset of peers with which each unit interacts is known as the \textit{reference group} \citep{Manski_2013_EJ}. $\square$
\end{example}

\begin{example}[Non-exchangeable peers]
\upshape  The case in which $h_0(\mathbf{d}_g)=\mathbf{d}_g$ does not provide any dimensionality reduction, as the only restriction it imposes is the existence of a known ordering between peers. This ordering is required to determine who is unit $i$'s nearest neighbor, second nearest neighbor and so on. Such an ordering can be based for example on geographic distance, a spatial weights matrix as used in spatial econometrics, frequency of interaction on social media, etc. $\square$
\end{example}

In what follows, $\mathbf{0}_ g$ and $\mathbf{1}_g$ will denote $n_g$-dimensional vectors of zeros and ones, respectively. Throughout the paper, I will assume that all the required moments of the potential outcomes are bounded. Unit-level direct effects are defined as differences in potential outcomes switching own treatment assignment for a fixed peer assignment $\mathbf{h}_0$, $Y_{ig}(1,\mathbf{h}_0)-Y_{ig}(0,\mathbf{h}_0)$. Unit-level spillover effects are defined as differences in potential outcomes switching peer assignments for a fixed own assignment $d$, $Y_{ig}(d,\mathbf{h}_0)-Y_{ig}(d,\mathbf{\tilde{h}}_0)$.

Given a vector of observed assignments $(D_{ig},\mathbf{D}_{(i)g})$, the observed outcome is given by $Y_{ig}(D_{ig},h_0(\mathbf{D}_{(i)g}))$ and can be written as:

\[Y_{ig}=\sum_{d\in\{0,1\}}\sum_{\mathbf{h}_0\in\mathcal{H}^0} Y_{ig}(d,\mathbf{h}_0)\1(D_{ig}=d)\1(h_0(\mathbf{D}_{(i)g})=\mathbf{h}_0).\]

To fix ideas, consider a household with three children, $n_g+1=3$. In this household, each kid has two siblings, with assignments $d_1$ and $d_2$, so $\mathbf{d}_g=(d_1,d_2)$. If the true treatment rule $h_0(\cdot)$ is the identity function, the potential outcome has the form $Y_{ig}(d,d_1,d_2)$ and hence each unit can have up to $2^{(n_g+1)}=8$ different potential outcomes. In this case, $Y_{ig}(1,0,0)-Y_{ig}(0,0,0)$ is the direct effect of the treatment when both of unit $i$'s siblings are untreated, $Y_{ig}(0,1,0)-Y_{ig}(0,0,0)$ is the spillover effect on unit $i$ of treating unit $i$'s first sibling, and so on. The average effect of assignment $(d,d_1,d_2)$ compared to $(\tilde{d},\tilde{d}_1,\tilde{d}_2)$ is thus given by $\E[Y_{ig}(d,d_1,d_2)]-\E[Y_{ig}(\tilde{d},\tilde{d}_1,\tilde{d}_2)]$. On the other hand, under an exchangeable treatment rule, the potential outcome can be written as $Y_{ig}(d,s)$ where $s=0,1,2$ is the number of treated siblings. The total number of different potential outcomes is $2(n_g+1)=6$, so exchangeability reduces the dimensionality of the effective assignments set from exponential to linear in group size. 

I assume perfect compliance, which means that all units receive the treatment they are assigned to. I analyze the case of imperfect compliance in \cite{Vazquez-Bare_2020_IV}. The data come from an infinite population of groups for which the researcher observes the outcomes at the unit level and the vector of treatment assignments in each group. Furthermore, by virtue of random assignment, potential outcomes are independent of treatment assignment. I formalize these features as follows.

\begin{assumption}[Sampling and random assignment]\label{assu:indep} Let $\mathbf{y}_{ig}=(Y_{ig}(d,\mathbf{h}_0))_{(d,\mathbf{h}_0)\in\{0,1\}\times \mathcal{H}_0}$ be the vector of potential outcomes for each unit $i$ in group $g$ and $\mathbf{y}_g=(\mathbf{y}_{1g}',\ldots,\mathbf{y}_{n_g+1,g}')'$.
\begin{enumerate}[(a)]
\item The vectors $(\mathbf{y}_1',\mathbf{D}_1')',\ldots,(\mathbf{y}_G',\mathbf{D}_G')'$ are sampled independently from an infinite population.
\item For any $g$ and $l$ such that $n_g=n_l$, $(\mathbf{y}_g',\mathbf{D}_g')'$ and $(\mathbf{y}_l',\mathbf{D}_l')'$ are identically distributed.
\item For each $g$, the elements in $\mathbf{y}_g$ are identically distributed.
\item For each $g$, $\mathbf{y}_g\indep \mathbf{D}_g$.
\end{enumerate}
\end{assumption}

Parts (a) and (b) in Assumption \ref{assu:indep} state that we observe a random sample of independent groups, and all groups with the same size are identically distributed. Part (c) indicates that units are identically distributed within each group, so that, for instance, $\E[Y_{ig}(d,\mathbf{h}_0)]$ is not indexed by $i$ (or $g$, given part (b)), and the same holds for other moments of the potential outcomes. Note that this condition does not prevent average potential outcomes (or other moments) to differ conditional on covariates. For example, if $X_{ig}$ denotes gender, the setup allows for $\E[Y_{ig}(d,\mathbf{h}_0)|X_{ig}=male]\ne \E[Y_{ig}(d,\mathbf{h}_0)|X_{ig}=female]$. Additionally, the identification results in the paper can be adapted to the case of non-identical distributions within group by switching focus from $\E[Y_{ig}(d,\mathbf{h}_0)]$ to $\sum_{i=1}^{n_g+1}\E[Y_{ig}(d,\mathbf{h}_0)]/(n_g+1)$. Finally, part (d) states that the treatment is randomly assigned and hence the vector of treatment indicators is independent of potential outcomes.

In practice, the true $h_0(\cdot)$ is usually unknown, and the researcher needs to posit a candidate $h(\cdot)$ that may or may not coincide with $h_0(\cdot)$. Given the lack of knowledge on the true assignment, a function $h(\cdot)$ that imposes fewer restrictions on the potential outcomes has a lower risk of misspecification. To formalize this idea, I introduce the following definition.

\begin{definition}[Coarseness]\label{def:coarseness}
Given two treatment rules $h(\cdot):\mathcal{D}\to \mathcal{H}$ and $\tilde{h}(\cdot):\mathcal{D}\to \tilde{\mathcal{H}}_g$, we say $h(\cdot)$ is coarser than $\tilde{h}(\cdot)$ if there exists another function $f(\cdot):\tilde{\mathcal{H}}_g\to \mathcal{H}$ such that $h(\mathbf{d}_g)=f\circ \tilde{h}(\mathbf{d}_g)$ for all $\mathbf{d}_g\in\mathcal{D}$.
\end{definition}

Intuitively, this means that $h(\cdot)$ gives a ``cruder'' summary of $\mathbf{d}_g$ (i.e. it discards more information) than $\tilde{h}(\cdot)$. In other words, a coarser function imposes more restrictions on the potential outcomes. For example, the exchangeable assignment $h(\mathbf{d}_g)=\mathbf{1}_g'\mathbf{d}_g$ is coarser than the identity function $\tilde{h}(\mathbf{d}_g)=\mathbf{d}_g$, and the reference group assignment $h(\mathbf{d}_g)=(d_1,d_2)$ is coarser than $\tilde{h}(\mathbf{d}_g)=(d_1,d_2,d_3,d_4)$. 

The next section addresses identification of average potential outcomes when the true treatment rule $h_0(\cdot)$ is possibly unknown.


\section{Identification}\label{sec:identification}

In what follows, let $\mathbf{H}_{ig}=h(\mathbf{D}_{(i)g})$ be the observed value of the chosen treatment rule, and let $\mathbf{H}^0_{ig}=h_0(\mathbf{D}_{(i)g})$. The following  result links observed outcomes, potential outcomes and effective treatment assignments, and will be used in the upcoming theorems.

\begin{lemma}[Nonparametric Identification]\label{lemma:identif}
Suppose Assumption \ref{assu:indep} holds and let $h_0(\cdot):\mathcal{D} \to \mathcal{H}_0$ be the true treatment rule. Given a treatment rule $h(\cdot):\mathcal{D} \to \mathcal{H}$, for any pair $(d,\mathbf{h})\in\{0,1\}\times\mathcal{H}$ such that $\P[D_{ig}=d,\mathbf{H}_{ig}=\mathbf{h}]>0$ and for any measurable function $m(\cdot)$,
\[\E[m(Y_{ig})|D_{ig}=d,\mathbf{H}_{ig}=\mathbf{h}]=\sum_{\mathbf{h}_0\in\mathcal{H}_0}\E[m(Y_{ig}(d,\mathbf{h}_0))]\P[\mathbf{H}^0_{ig}=\mathbf{h}_0|D_{ig}=d,\mathbf{H}_{ig}=\mathbf{h}].\]
In particular, if $h_0(\cdot)$ is coarser than $h(\cdot)$, then
\[\E[m(Y_{ig})|D_{ig}=d,\mathbf{H}_{ig}=\mathbf{h}]=\E[m(Y_{ig}(d,h_0(\mathbf{h})))].\]
\end{lemma}

Different choices of the function $m(\cdot)$ lead to different estimands of interest. For example, setting $m(\cdot)=\1(\cdot\le y)$ for some $y\in\mathbbm{R}$ yields $\E[m(Y_{ig})|D_{ig}=d,\mathbf{H}_{ig}=\mathbf{h}]=F_Y(y|D_{ig}=d,\mathbf{H}_{ig}=\mathbf{h})$ where $F_Y(\cdot|\cdot)$ is the conditional cdf of $Y_{ig}$. This choice of $m(\cdot)$ can be used to identify the distribution of potential outcomes. In what follows, unless explicitly stated, I will let $m(\cdot)$ be the identity function to reduce notation.

Lemma \ref{lemma:identif} shows that the average observed outcome among units facing $D_{ig}=d$ and $\mathbf{H}_{ig}=\mathbf{h}$ averages the potential outcomes over all the assignments $\mathbf{h}_0$ that are consistent with $(D_{ig},\mathbf{H}_{ig})=(d,\mathbf{h})$, as long as the probability of $(d,\mathbf{h})$ is not zero. 

To illustrate Lemma \ref{lemma:identif}, consider the previous example with three units and where $h_0(\cdot)$ is the identity function so the potential outcome has the form $Y_{ig}(d,d_1,d_2)$. Suppose we posit an exchangeable treatment rule $h(\mathbf{d}_g)=\mathbf{1}'_g\mathbf{d}_g$ and thus $\mathbf{H}_{ig}=S_{ig}=\sum_{j\ne i} D_{jg}$ which is a scalar counting how many of unit $i$'s peers are treated. By Lemma \ref{lemma:identif}, if $\P[D_{ig}=0,S_{ig}=1]>0$, $\E[Y_{ig}|D_{ig}=0,S_{ig}=1]$ equals a weighted average of $\E[Y_{ig}(0,1,0)]$ and $\E[Y_{ig}(0,0,1)]$, with weights given by the conditional probabilities of these different assignments. 

In general, $(D_{ig},\mathbf{H}_{ig})=(d,\mathbf{h})$ may be consistent with many different effective assignments $\mathbf{h}_0$. When $h_0(\cdot)$ is coarser than $h(\cdot)$, however, the value of $\mathbf{h}_0$ is uniquely determined. In such cases, the second part of Lemma \ref{lemma:identif} shows that $\E[Y_{ig}|D_{ig}=d,\mathbf{H}_{ig}=\mathbf{h}]$ identifies the value of the average potential outcome consistent with that assignment. For example, in the case of $n_g=2$, suppose that the true $h_0(\cdot)$ is exchangeable, so that outcomes have the form $Y_{ig}(d,s)$ with $s=0,1,2$, and suppose we posit $h(\mathbf{d}_g)=\mathbf{d}_g=(d_1,d_2)$. Setting $\mathbf{H}_{ig}=(1,1)$ implies that the sum of treated peers is equal to 2, and therefore $\E[Y_{ig}|D_{ig}=0,\mathbf{H}_{ig}=(1,1)]=\E[Y_{ig}(0,2)]$. In particular, this result implies that if $h(\cdot)$ is equal to $h_0(\cdot)$, $\E[Y_{ig}|D_{ig}=d,\mathbf{H}_{ig}=\mathbf{h}]=\E[Y_{ig}(d,\mathbf{h})]$.

\begin{remark}[Implications for $h_0(\cdot)$]
\upshape When $h_0(\cdot)$ is coarser than $h(\cdot)$, Lemma \ref{lemma:identif} implies restrictions on the shape of the true treatment rule $h_0(\cdot)$. More precisely, if there exists an $m(\cdot)$ and a pair $(\mathbf{h},\mathbf{\tilde{h}})$ such that $\E[m(Y_{ig})|D_{ig}=d,\mathbf{H}_{ig}=\mathbf{h}]\ne \E[m(Y_{ig})|D_{ig}=d,\mathbf{H}_{ig}=\mathbf{\tilde{h}}]$, then it follows that $h_0(\mathbf{h})\ne h_0(\mathbf{\tilde{h}})$.\footnote{To see this, note that $\E[m(Y_{ig})|D_{ig}=d,\mathbf{H}_{ig}=\mathbf{h}]\ne \E[m(Y_{ig})|D_{ig}=d,\mathbf{H}_{ig}=\mathbf{\tilde{h}}]$ implies $\E[m(Y_{ig}(d,h_0(\mathbf{h})))]\ne \E[m(Y_{ig}(d,h_0(\mathbf{\tilde{h}})))]$ by Lemma \ref{lemma:identif}. If $h_0(\mathbf{h})=h_0(\mathbf{\tilde{h}})$, then $\E[m(Y_{ig}(d,h_0(\mathbf{h})))]\ne \E[m(Y_{ig}(d,h_0(\mathbf{h})))]$ which gives a contradiction, and thus $h_0(\mathbf{h})\ne h_0(\mathbf{\tilde{h}})$.} For example, suppose each unit has two peers and that $h(\mathbf{d}_g)=\mathbf{d_g}=(d_1,d_2)$. In this case, if $\E[m(Y_{ig})|D_{ig}=d,\mathbf{H}_{ig}=(1,0)]\ne \E[m(Y_{ig})|D_{ig}=d,\mathbf{H}_{ig}=(0,1)]$, then $h_0(1,0)\ne h_0(0,1)$ which rules out, for instance, an exchangeable treatment rule $h_0(d_1,d_2)=d_1+d_2$ and an ``interaction'' treatment rule $h_0(d_1,d_2)=d_1\cdot d_2$. On the other hand, finding that $\E[m(Y_{ig})|D_{ig}=d,\mathbf{H}_{ig}=\mathbf{h}]= \E[m(Y_{ig})|D_{ig}=d,\mathbf{H}_{ig}=\mathbf{\tilde{h}}]$ for some $m(\cdot)$ does not imply that $h_0(\mathbf{h})= h_0(\mathbf{\tilde{h}})$, since the equality between moments could fail for a different choice of $m(\cdot)$. $\square$
\end{remark}

\begin{remark}[Pooled estimands]\label{rmk:pooled}
\upshape Coarse treatment rules can be used not only as a modeling assumption on potential outcomes but also as summary measures of average potential outcomes and treatment effects. For instance, setting $h(\cdot)$ equal to a constant function, which ignores $\mathbf{d}_g$, averages over all possible peers' assignments: $\E[Y_{ig}|D_{ig}=d]=\sum_{\mathbf{h}_0\in\mathcal{H}_0}\E[Y_{ig}(d,\mathbf{h}_0)]\P[\mathbf{H}^0_{ig}=\mathbf{h}_0|D_{ig}=d]$. Notice that $\E[Y_{ig}|D_{ig}=1]-\E[Y_{ig}|D_{ig}=0]=\sum_{\mathbf{h}_0\in\mathcal{H}_0}\E[Y_{ig}(1,\mathbf{h}_0)]\P[\mathbf{H}^0_{ig}=\mathbf{h}_0|D_{ig}=1]-\sum_{\mathbf{h}_0\in\mathcal{H}_0}\E[Y_{ig}(0,\mathbf{h}_0)]\P[\mathbf{H}^0_{ig}=\mathbf{h}_0|D_{ig}=0]$, which is the the super-population analog of the direct average causal effect defined by \citet{Hudgens-Halloran_2008_JASA}. Alternatively, let  $s=\mathbf{1}'_g\mathbf{d}_g$, and define $h(\mathbf{d}_g)=\1(s>0)$ which equals one if there is at least one treated peer. Let $S_{ig}=\sum_{j\ne i} D_{jg}$ be the observed number of treated peers for unit $i$. Then, by Lemma \ref{lemma:identif}, $\E[Y_{ig}|D_{ig}=d,S_{ig}>0]=\sum_{\mathbf{h}_0\in\mathcal{H}_0} \E[Y_{ig}(d,\mathbf{h}_0)]\P[\mathbf{H}^0_{ig}=\mathbf{h}_0|D_{ig}=d,S_{ig}>0]$. Consider the difference between untreated units with at least one treated peer and untreated units with no treated peers:
\[\Delta=\E[Y_{ig}|D_{ig}=0,S_{ig}>0]-\E[Y_{ig}|D_{ig}=0,S_{ig}=0].\]
Then, given that $S_{ig}=0$ implies that $\mathbf{D}_{(i)g}=\mathbf{0}_g$, we have that:
\[\Delta=\sum_{\mathbf{h}_0\in\mathcal{H}_0} \E[Y_{ig}(0,\mathbf{h}_0)-Y_{ig}(0,\mathbf{0})]\P[\mathbf{H}^0_{ig}=\mathbf{h}_0|D_{ig}=0,S_{ig}>0]\]
where $\mathbf{0}=h_0(\mathbf{0}_g)$. Thus, $\Delta$ recovers a weighted average of spillover effects on untreated units weighted by the probabilities of the different assignments, which is analogous to the average indirect causal effect of \citet{Hudgens-Halloran_2008_JASA} but for a general assignment mechanism. A natural generalization of this idea is to split $S_{ig}$ into categories such as $S_{ig}=0$, $1\le S_{ig}\le k$, $k+1\le S_{ig}\le n_g$ and so on. Section \ref{sec:empapp} illustrates how to estimate these pooled parameters using a saturated regression. $\square$
\end{remark}

\begin{remark}[Partial population experiments]\label{rmk:ppexp}
\upshape A popular design when analyzing spillover effects is the partial population design \citep{Moffit_2001,Duflo-Saez_2003_QJE,Hirano-Hahn_2010_EL,Baird-etal_2018_Restat}. In its simplest form, groups are randomly divided into treated and controls based on a binary indicator $T_g$. Then, within the groups with $T_g=1$, treatment $D_{ig}$ is randomly assigned at the individual level. In these type of experiments, spillover effects are often estimated as the average difference between control units in treated groups and control units in pure control groups,
\begin{equation*}
\Delta_{\mathsf{PP}} = \E[Y_{ig}|D_{ig}=0,T_g=1]-\E[Y_{ig}|T_g=0].
\end{equation*}
Redefining the vector of treatment assignments as $(D_{ig},\mathbf{D}_{(i)g},T_g)=(d,\mathbf{d}_g,t)$ and setting $h(\mathbf{d}_g,t)=t$, if $(\mathbf{D}_g,T_g)$ is independent of potential outcomes, Lemma \ref{lemma:identif} implies that:
\[\Delta_{\mathsf{PP}}=\sum_{\mathbf{h}_0\in\mathcal{H}_0} \E[Y_{ig}(0,\mathbf{h}_0)-Y_{ig}(0,\mathbf{0})]\P[\mathbf{H}^0_{ig}=\mathbf{h}_0|D_{ig}=0,T_g=1]\]
which averages over all the possible number of treated peers that an untreated unit can have in a treated group.  The generalization to experiments with more than two categories \citep[see e.g.][]{Crepon-etal_2013_QJE} is straightforward. $\square$
\end{remark}


\subsection{Difference in means}\label{sec:dm}

The difference in means estimand $\beta_\mathsf{D}=\E[Y_{ig}|D_{ig}=1]-\E[Y_{ig}|D_{ig}=0]$, which compares the average observed outcomes between treated and controls, is arguably the most common estimand when analyzing randomized experiments. It is well known that, in the absence of spillovers, $\beta_{\mathsf{D}}$ equals the average treatment effect (ATE) when the treatment is randomly assigned. An estimate for the ATE can be calculated by estimating the model:
\begin{equation}\label{eq:diffmeans}
Y_{ig}=\alpha_{\mathsf{D}}+\beta_{\mathsf{D}} D_{ig}+u_{ig}, \quad \E[u_{ig}]=\cov(D_{ig},u_{ig})=0.
\end{equation}
The following results characterizes the difference in means $\beta_{\mathsf{D}}$ in the presence of spillovers. In what follows, let $\mathbf{0}=h_0(\mathbf{0}_g)$.
\begin{theorem}[Difference in means]\label{thm:dm}
Under Assumption \ref{assu:indep}, the coefficient $\beta_{\mathsf{D}}$ from Equation \eqref{eq:diffmeans} can be written as:
\begin{align*}
\beta_{\mathsf{D}}=\E[Y_{ig}(1,\mathbf{0})-Y_{ig}(0,\mathbf{0})]&+\sum_{\mathbf{h}_0\in\mathcal{H}_0}\E[Y_{ig}(1,\mathbf{h}_0)-Y_{ig}(1,\mathbf{0})]\P[\mathbf{H}^0_{ig}=\mathbf{h}_0|D_{ig}=1]\\
&-\sum_{\mathbf{h}_0\in\mathcal{H}_0}\E[Y_{ig}(0,\mathbf{h}_0)-Y_{ig}(0,\mathbf{0})]\P[\mathbf{H}^0_{ig}=\mathbf{h}_0|D_{ig}=0].
\end{align*}
\end{theorem}
Hence, the difference-in-means estimand equals the direct effect without treated peers $\E[Y_{ig}(1,\mathbf{0})-Y_{ig}(0,\mathbf{0})]$ plus the difference in weighted averages of spillover effects under treatment and under control. In general, the sign of this difference is undetermined, as it depends on the relative magnitudes of the average spillover effects on treated and controls. Thus, the difference in means can be larger than, smaller than or equal to the average direct effect $\E[Y_{ig}(1,\mathbf{0})-Y_{ig}(0,\mathbf{0})]$. In particular, if the spillover effects on treated units are equal to zero and the spillover effects on controls are of the same sign that the direct effect, the difference in means will underestimate the average direct effect without treated peers. This case matches the commonly invoked intuition that spillovers ``contaminate'' the control group.


\subsection{Reduced-form linear-in-means regression}\label{sec:lim}

Equation \eqref{eq:diffmeans} may give an incomplete assessment of the effect of a treatment because it completely ignores the presence of spillovers. When trying to explicitly estimate spillover effects, a common strategy is to estimate a reduced-form linear-in-means (RF-LIM) regression, which is given by:
\begin{equation}\label{eq:lim}
Y_{ig}=\alpha_{\ell}+\beta_{\ell}D_{ig}+\gamma_{\ell}\bar{D}^{(i)}_g+\eta_{ig},\qquad \E[\eta_{ig}]=\cov(D_{ig},\eta_{ig})=\cov(\bar{D}^{(i)}_g,\eta_{ig})=0
\end{equation}
where
\[\bar{D}^{(i)}_g=\frac{1}{n_g}\sum_{j\ne i}D_{jg}.\]
This is a regression of the outcome on a treatment indicator and the proportion of treated peers. In this specification, $\beta_\ell$ intends to capture a direct effect whereas $\gamma_{\ell}$ is seen as a measure of spillover effects, since it captures the average change in outcomes in response to a change in the proportion of treated neighbors. While the parameters $(\beta_\ell,\gamma_\ell)$ can be interpreted as linear projection coefficients, the following result shows that they do not have a causal interpretation in general. 
\begin{theorem}[RF-LIM regression]\label{thm:lim}
Under Assumption \ref{assu:indep}, the coefficients $(\beta_\ell,\gamma_\ell)$ from Equation \eqref{eq:lim} can be written as:
\begin{align*}
\beta_\ell&=\E[Y_{ig}|D_{ig}=1]-\E[Y_{ig}|D_{ig}=0]-\frac{\gamma_\ell}{n_g}(\E[S_{ig}|D_{ig}=1]-\E[S_{ig}|D_{ig}=0])\\
\gamma_\ell&=\sum_{s=1}^{n_g}\phi_0(s)(\E[Y_{ig}|D_{ig}=0,S_{ig}=s]-\E[Y_{ig}|D_{ig}=0,S_{ig}=0])\\
&+\sum_{s=1}^{n_g}\phi_1(s)(\E[Y_{ig}|D_{ig}=1,S_{ig}=s]-\E[Y_{ig}|D_{ig}=1,S_{ig}=0])
\end{align*}
where for $d=0,1$,
\[\phi_d(s)=\frac{n_g\P[D_{ig}=d]\P[S_{ig}=s|D_{ig}=d]}{\P[D_{ig}=0]\V[S_{ig}|D_{ig}=0]+\P[D_{ig}=1]\V[S_{ig}|D_{ig}=1]}\cdot(s-\E[S_{ig}|D_{ig}=d]).\]
\end{theorem}
Theorem \ref{thm:lim} characterizes the coefficients from the linear projection of $Y_{ig}$ into $D_{ig}$ and $\bar{D}^{(i)}_g$. The coefficient $\beta_\ell$ equals the difference in means minus an adjustment factor that depends on $\gamma_\ell$ and the relationship between treatment assignments within group. This parameter is similar to the one analyzed in Theorem \ref{thm:dm} and hence does not have a direct causal interpretation in general. 

The coefficient $\gamma_\ell$ equals a linear combination of differences $\E[Y_{ig}|D_{ig}=d,S_{ig}=s]-\E[Y_{ig}|D_{ig}=d,S_{ig}=0]$ across all values of $s$ with weights $(\phi_0(s),\phi_1(s))_s$. Two factors obscure the causal interpretation of this coefficient. On the one hand, the magnitudes $\E[Y_{ig}|D_{ig}=d,S_{ig}=s]$ implicitly impose an exchangeable treatment rule that may be misspecified in general. The interpretation of these expectations is given in Lemma \ref{lemma:identif}. On the other hand, even if this treatment rule was correctly specified, so that $\E[Y_{ig}|D_{ig}=d,S_{ig}=s]-\E[Y_{ig}|D_{ig}=d,S_{ig}=0]=\E[Y_{ig}(d,s)-Y_{ig}(d,0)]$, these effects are combined using weights that can be negative. Indeed, it can be seen that $\sum_{s=0}^{n_g}\phi_d(s)=0$ for $d=0,1$ and hence some of the weights are necessarily negative (specifically, the ones corresponding to low values of $s$). Thus, $\gamma_\ell$ may be zero or negative when all the average spillover effects are positive or vice versa. 

To illustrate the importance of these issues in an empirical setting, Section \ref{sec:empapp} shows a case in which the estimates of $\beta_\ell$ and $\gamma_\ell$ are close to zero and not statistically significant, even when the estimated average spillover effects are all large and statistically significant when estimated nonparametrically. 

The following result provides conditions under which the coefficients from a RF-LIM regression have a clear causal interpretation, that is, they can be written as proper weighted averages of direct and/or spillover effects.
\begin{corollary}[Correctly-specified RF-LIM regression]\label{coro:lim}
Suppose that, in addition to Assumption \ref{assu:indep}, the following conditions hold:
\begin{enumerate}[(i)]
\item Exchangeability: $Y_{ig}(d,h_0(\mathbf{d}_g))=Y_{ig}(d,s)$ where $s=\mathbf{1}'_g\mathbf{d}_g$,
\item Linearity: for each $d=0,1$ there is a constant $\kappa_d$ such that $\E[Y_{ig}(d,s)-Y_{ig}(d,s-1)]=\kappa_d$ for all $s\ge 1$.
\end{enumerate}
Then, the coefficients $(\beta_\ell,\gamma_\ell)$ from Equation \eqref{eq:lim} are:
\begin{align*}
\beta_\ell&=\E[Y_{ig}(1,0)-Y_{ig}(0,0)]+(\kappa_1-\kappa_0)\{(1-\lambda)\E[S_{ig}|D_{ig}=1]+\lambda\E[S_{ig}|D_{ig}=0]\}\\
\gamma_\ell&=\lambda\E[Y_{ig}(1,n_g)-Y_{ig}(1,0)]+(1-\lambda)\E[Y_{ig}(0,n_g)-Y_{ig}(0,0)]
\end{align*}
where
\[\lambda=\frac{\P[D_{ig}=1]\V[S_{ig}|D_{ig}=1]}{\P[D_{ig}=1]\V[S_{ig}|D_{ig}=1]+\P[D_{ig}=0]\V[S_{ig}|D_{ig}=0]}\in(0,1).\]
\end{corollary}
The above result highlights two restrictions that the RF-LIM regression implicitly imposes on potential outcomes: (i) peers are exchangeable, so potential outcomes only depend on own treatment and the number of treated peers, and (ii) average spillover effects are linear in $s$ so that, for instance, the effect of having two treated peers is twice as large as the effect of having one treated peer. If these conditions hold, $\gamma_\ell$ recovers a weighted average of the effects of having all treated peers for treated and untreated units, where the weights are positive and sum to one. Hence, the RF-LIM regression is robust to some heterogeneity in spillover effects $\E[Y_{ig}(d,s)-Y_{ig}(d,0)]$ both over $s$ and over $d$, but suffers from potentially severe misspecification when spillover effects are nonlinear. 

On the other hand, $\beta_\ell$ does not recover a causal effect in general. In the particular case in which $\kappa_1=\kappa_0$, that is, the average spillover effects are equal for treated and untreated units, $\beta_\ell$ equals the average direct effect with no treated peers $\E[Y_{ig}(1,0)-Y_{ig}(0,0)]$.

\begin{remark}[Structural LIM models]
\upshape Structural LIM models include average peers outcomes $\bar{Y}_{g}^{(i)}$ on the right-hand side of an equation like \eqref{eq:lim} \citep{Manski_1993_Restud,Kline-Tamer_2019}. Under this specification, in addition to the dependence on peers' treatments, a unit's outcome can be affected by peers' outcomes, a phenomenon known as \textit{endogenous effects} \citep{Manski_1993_Restud} or \textit{interference by contagion} \citep{Ogburn-VanderWeele_2014_SS}. While I do not consider identification of endogenous effects in this paper, Section \ref{app:structural} in the supplemental appendix shows that Equation \eqref{eq:lim} can be rationalized as the reduced form of the structural LIM model:
\[Y_{ig}=\phi(D_{ig},\mathbf{D}_{(i)g})+\gamma \bar{Y}_{g}^{(i)}+u_{ig}\]
under the assumptions of Corollary \ref{coro:lim}. $\square$
\end{remark}

A straightforward way to make Equation \eqref{eq:lim} more flexible is to include an interaction term between own treatment indicator and the proportion of treated peers:
\begin{equation}\label{eq:lim_inter}
Y_{ig}=\tilde{\alpha}_{\ell}+\tilde{\beta}_{\ell}D_{ig}+\gamma^0_{\ell}\bar{D}^{(i)}_g(1-D_{ig})+\gamma^1_{\ell}\bar{D}^{(i)}_gD_{ig}+\xi_{ig}
\end{equation}
where $\E[\xi_{ig}]=\cov(D_{ig},\xi_{ig})=\cov(\bar{D}^{(i)}_g(1-D_{ig}),\xi_{ig})=\cov(\bar{D}^{(i)}_gD_{ig},\xi_{ig})=0$. The following result characterizes the coefficients from this specification.
\begin{theorem}[Interacted RF-LIM regression]\label{thm:lim_inter}
Under Assumption \ref{assu:indep}, the coefficients $(\tilde{\beta}_\ell,\gamma^0_\ell,\gamma^1_\ell)$ from Equation \eqref{eq:lim_inter} can be written as:
\begin{align*}
\tilde{\beta}_\ell&=\E[Y_{ig}|D_{ig}=1]-\E[Y_{ig}|D_{ig}=0]-\left(\frac{\gamma^1_\ell}{n_g}\E[S_{ig}|D_{ig}=1]-\frac{\gamma^0_\ell}{n_g}\E[S_{ig}|D_{ig}=0]\right)\\
\gamma^0_\ell&=\sum_{s=1}^{n_g}\omega_0(s)(\E[Y_{ig}|D_{ig}=0,S_{ig}=s]-\E[Y_{ig}|D_{ig}=0,S_{ig}=0])\\
\gamma^1_\ell&=\sum_{s=1}^{n_g}\omega_1(s)(\E[Y_{ig}|D_{ig}=1,S_{ig}=s]-\E[Y_{ig}|D_{ig}=1,S_{ig}=0])
\end{align*}
where for $d=0,1$,
\[\omega_d(s)=\frac{n_g\P[S_{ig}=s|D_{ig}=d]}{\V[S_{ig}|D_{ig}=d]}\cdot(s-\E[S_{ig}|D_{ig}=d]).\]
\end{theorem}
According to this theorem, an interacted RF-LIM regression separates the spillover components $\E[Y_{ig}|D_{ig}=0,S_{ig}=s]-\E[Y_{ig}|D_{ig}=0,S_{ig}=0]$ and $\E[Y_{ig}|D_{ig}=1,S_{ig}=s]-\E[Y_{ig}|D_{ig}=1,S_{ig}=0]$. However, the issue of negative weights remains for each component, since $\sum_{s=0}^{n_g}\omega_d(s)=0$ for $d=0,1$ and hence some of the weights are necessarily negative.

Finally, the following result shows that when peers are exchangeable and spillover effects are linear, the interacted RF-LIM regression can recover all the causal parameters of interest. 

\begin{corollary}[Correctly-specified interacted RF-LIM regression]\label{coro:lim_inter}
Suppose that, in addition to Assumption \ref{assu:indep}, the following conditions hold:
\begin{enumerate}[(i)]
\item Exchangeability: $Y_{ig}(d,h_0(\mathbf{d}_g))=Y_{ig}(d,s)$ where $s=\mathbf{1}'_g\mathbf{d}_g$,
\item Linearity: for each $d=0,1$ there is a constant $\kappa_d$ such that $\E[Y_{ig}(d,s)-Y_{ig}(d,s-1)]=\kappa_d$ for all $s\ge 1$.
\end{enumerate}
Then, the coefficients $(\tilde{\beta}_\ell,\gamma^0_\ell,\gamma^1_\ell)$ from Equation \eqref{eq:lim_inter} can be written as:
\begin{align*}
\tilde{\beta}_\ell&=\E[Y_{ig}(1,0)-Y_{ig}(0,0)]\\
\gamma^0_\ell&=\E[Y_{ig}(0,n_g)-Y_{ig}(0,0)]\\
\gamma^1_\ell&=\E[Y_{ig}(1,n_g)-Y_{ig}(1,0)].
\end{align*}
\end{corollary}
According to this result, the coefficients from a correctly-specified interacted RF-LIM regression recover the average direct effect without treated peers and the spillover effects of having all peers treated, for treated and untreated units separately. Because of linearity, all the remaining spillover effects can be recovered by appropriately rescaling $\gamma^d_\ell$. For example, the average spillover effect from having one treated peer on an untreated unit is $\E[Y_{ig}(0,1)-Y_{ig}(0,0)]=\gamma^0_\ell/n_g$.


\section{Estimation and inference}\label{sec:estimation}

The previous sections provide conditions under which average direct and spillover effects can be nonparametrically identified by exploiting variation over own and peers' assignments. Because these magnitudes can be written as population averages, it is straightforward to construct their corresponding estimators based on sample cell means. The main challenge for estimation arises when groups are large. A large number of units per group (as in households with multiple family members or classrooms with a large number of students) requires estimating a large number of means in each of the cells defined by the treatment assignments. In such cases, the probability of observing some assignments can be close to zero and the number of observations in each cell may be too small to estimate the average potential outcomes. 

For example, suppose the treatment is assigned as an independent coin flip with probability $p=1/2$. Under this assignment we would expect most groups to have about half its units treated, so when groups have, say, 10 units, 5 of them would be treated on average. The probability of observing groups with zero or all treated units, on the other hand, will be close to zero, and thus the average potential outcomes corresponding to these ``tail assignments'' will be hard to estimate precisely.

So far, the analysis has been done taking group size as fixed. When group size is fixed, small cells are a finite sample problem that disappears as the sample grows. To account for this phenomenon asymptotically, in this section I will generalize this setting and consider double-array asymptotics in which the group size is allowed to grow with the sample size. The goal is to answer the question of how large groups can be relative to the total sample size to allow for valid estimation and inference. The key issue to obtain consistency and asymptotic normality will be to ensure that the number of observations in all cells grows sufficiently fast as the sample size increases. This setup is not intended to model a population in which groups are effectively infinitely large, but as a statistical device to approximate the distribution of estimators in finite samples when the number of parameters can be ``moderately'' large, in a sense that will be made more precise in this section. The case with fixed group size is a particular case of this setting. 

In this section I will assume that groups are equally sized, so that $n_g=n$. Recall that given a candidate treatment rule $h(\cdot)$ and $\mathbf{h}=h(\mathbf{d}_g)$, the effective treatment assignments  are given by $(d,\mathbf{h}_g)\in\{0,1\}\times\mathcal{H}$. As formalized in Assumption \ref{assu:sampling} below, $h(\cdot)$ is not assumed to equal the true treatment rule, but the true treatment rule $h_0(\cdot)$ has to be coarser than $h(\cdot)$ as specified in Definition \ref{def:coarseness}. To make the notation more compact, I will let $\mathcal{A}_n=\{0,1\}\times\mathcal{H}$ where the notation makes the dependence of this set on the group size explicit. The cardinality of this set is denoted by $|\mathcal{A}_n|$, which indicates the total number of parameters to be estimated. The observed effective treatment assignments will be $(D_{ig},\mathbf{H}_{ig})=\mathbf{A}_{ig}$, taking values $\mathbf{a}\in\mathcal{A}_n$, and $\mu(\mathbf{a})=\E[Y_{ig}|\mathbf{A}_{ig}=\mathbf{a}]$.

Because $\mathbf{A}_{ig}$ takes on a finite number of values, all the conditional means can be estimated jointly through the regression:
\begin{align}
Y_{ig}=\alpha+\sum_{\mathbf{a}\in \mathcal{A}_{n,0}}\beta_\mathbf{a} \1(\mathbf{A}_{ig}=\mathbf{a})+\nu_{ig}, \quad \E[\nu_{ig}|\mathbf{A}_{ig}]=0
\end{align}
where $\mathcal{A}_{n,0}=\mathcal{A}_n\backslash \{\mathbf{a}_0\}$ and $\mathbf{a}_0$ is the baseline treatment assignment (typically, the assignment in which no unit is treated). Because this regression is fully saturated, by construction $\alpha=\E[Y_{ig}|\mathbf{A}_{ig}=\mathbf{a}_0]$ and $\beta_\mathbf{a}=\E[Y_{ig}|\mathbf{A}_{ig}=\mathbf{a}]-\E[Y_{ig}|\mathbf{A}_{ig}=\mathbf{a}_0]$. Hence, this regression can be seen as a nonparametric regression as it does not impose any functional form assumptions. Since all the coefficients are linear combinations of conditional means, it suffices to focus on the vector of means $\mu(\mathbf{a})$ to analyze the properties of the coefficient estimators.

Each treatment assignment mechanism determines a distribution $\pi(\cdot)$ over $\mathcal{A}_n$ where $\pi(\mathbf{a})=\P[\mathbf{A}_{ig}=\mathbf{a}]$ for $\mathbf{a}\in\mathcal{A}_n$. For example, when $\mathcal{A}_n=\{0,1\}$, if the treatment is assigned independently as a coin flip, $\pi(1)=\P[D_{ig}=1]=p$ and $\pi(0)=1-p$. Under the same assignment, with an exchangeable treatment rule, $\pi(d,s)=\P[D_{ig}=d,S_{ig}=s]={n \choose s}p^{s+d}(1-p)^{n+1-s-d}$. A key issue of this double-array asymptotic setup is that, since the size of the set $\mathcal{A}_n$ can increase with group size, the probabilities $\pi(\mathbf{a})$ can shrink towards zero for some (or all) assignments $\mathbf{a}\in\mathcal{A}_n$. The rate at which these probabilities decrease with the sample size is given by the experimental design. For instance, in the coin flip assignment just described, $\P[D_{ig}=0,S_{ig}=0]=(1-p)^{n+1}$ which decreases exponentially with $n$. Define:
\[\underline{\pi}_n=\min\limits_{\mathbf{a}\in\mathcal{A}_n} \pi(\mathbf{a})\]
which is the probability of the least likely treatment assignment. This probability, together with the total sample size, will determine the number of observations in the smallest assignment cell, that is, the number of observations available to estimate the ``hardest'' average potential outcome.

Let $\mathbf{A}_g=(\mathbf{A}_{1g},\ldots,\mathbf{A}_{n_g+1,g})$, $\mathbf{A}=(\mathbf{A}_1,\ldots,\mathbf{A}_G)$, and $\mathbf{Y}_g=(Y_{1g},Y_{2g},\ldots Y_{n_g+1,g})'$. I will assume the following.

\begin{assumption}[Sampling]\label{assu:sampling}
\item
\begin{enumerate}[(i)]
\item For $g=1,\ldots,G$, $(\mathbf{Y}_g',\mathbf{A}_g')$ are iid, and $n_g=n$.
\item The true treatment rule $h_0(\cdot)$ is coarser than $h(\cdot)$.
\item The potential outcomes are independent across $i$ within groups.
\end{enumerate}
\end{assumption}

Part (i) in Assumption \ref{assu:sampling} states that the researcher has access to a sample of $G$ independent and identically distributed equally-sized groups. When groups have different sizes (for example, households with 3, 4 or 5 siblings), the analysis can be performed separately for each group size. Section  \ref{app:group_size} of the supplemental appendix further discusses the case of unequally-sized groups. Part (ii)  allows the posited treatment rule $h(\cdot)$ to be different from the true treatment rule, but requires it to be flexible enough to break the dependence between $Y_{ig}$ and $\mathbf{A}_{jg}$ conditional on $\mathbf{A}_{ig}$ for $j\ne i$. Part (iii) assumes that potential outcomes are independent within groups, and hence the only source of dependence between the observed outcomes is the assignment $\mathbf{A}_g$. This condition can be relaxed to arbitrary dependence structures when the group size is fixed. Together, conditions (ii) and (iii) imply that observed outcomes are independent conditional on the assignments. Importantly, note that these conditions do not restrict the correlation between treatment assignments $\mathbf{A}_{ig}$ and $\mathbf{A}_{jg}$ in any way. In fact, the effective treatment assignments are correlated by construction, since $\mathbf{A}_{ig}$ depends on $D_{jg}$ and vice versa.

Given a sample of $G$ groups with $n+1$ units each, let $\1_{ig}(\mathbf{a})=\1(\mathbf{A}_{ig}=\mathbf{a})$, $N_g(\mathbf{a})=\sum_{i=1}^{n+1}\1_{ig}(\mathbf{a})$ and $N(\mathbf{a})=\sum_{g=1}^G N_g(\mathbf{a})$, so that $N_g(\mathbf{a})$ is the total number of observations receiving effective assignment $\mathbf{a}$ in group $g$ and $N(\mathbf{a})$ is the total number of observations receiving effective assignment $\mathbf{a}$ in the sample. The estimator for $\mu(\mathbf{a})$ is defined as:
\begin{align*}
\hat{\mu}(\mathbf{a})=\begin{cases}
\frac{\sum_{g=1}^G\sum_{i=1}^{n+1}Y_{ig}\1_{ig}(\mathbf{a})}{N(\mathbf{a})} \quad &\text{if } N(\mathbf{a})>0\\
\text{undefined} & \text{if } N(\mathbf{a})=0
\end{cases}
\end{align*}
Thus, the estimator for $\mu(\mathbf{a})$ is simply the sample average of the outcomes for observations receiving assignment $\mathbf{a}$, whenever there is at least one observation receiving this assignment.

The following assumption imposes some regularity conditions that are required for upcoming theorems. Let $\sigma^2(\mathbf{a})=\V[Y_{ig}|\mathbf{A}_{ig}=\mathbf{a}]$.
\begin{assumption}[Moments]\label{assu:moments}
There are constants $\underline{\sigma}$ and $b$ such that:
\[(i)\quad \inf\limits_{n}\min\limits_{\mathbf{a}\in\mathcal{A}_n}\sigma^2(\mathbf{a})\ge \underline{\sigma}^2>0,\qquad (ii) \quad \sup\limits_{n}\max\limits_{\mathbf{a}\in\mathcal{A}_n}\E[Y_{ig}^6|\mathbf{A}_{ig}=\mathbf{a}]\le b<\infty\]
\end{assumption}
Then we have the following result.
\begin{lemma}[Effective sample size]\label{lemma:sampsi}
Suppose Assumption \ref{assu:sampling}(i) holds, and consider an assignment mechanism $\pi(\cdot)$ such that $\pi(\mathbf{a})>0$ for all $\mathbf{a}\in\mathcal{A}_n$. If
\begin{equation}\label{eq:cond_cells}
\frac{\log |\mathcal{A}_n|}{G\underline{\pi}_n}\to 0
\end{equation}
then for any $c\in\mathbbm{R}$
\[\P\left[\min\limits_{\mathbf{a}\in\mathcal{A}_n}N(\mathbf{a})>c\right]\to 1.\]
\end{lemma}
Lemma \ref{lemma:sampsi} says that, under condition \eqref{eq:cond_cells}, the number of observations in the smallest cell will go to infinity, which implies that all the estimators are well defined asymptotically. Hence, condition \eqref{eq:cond_cells} formalizes the meaning of ``large sample'' in this context, and states that the number of groups has to be large relative to the total number of parameters and the probability of the least likely assignment. This expression can be interpreted as an invertibility condition for the design matrix of a linear regression model, in the specific case in which the regressors are mutually exclusive indicator variables. This requirement can be seen as a low-level condition that justifies the assumption of invertibility of the design matrix \citep[see e.g. Assumption 2 in][]{Cattaneo-Jansson-Newey_2018_JASA}. When this condition does not hold, small cells may result in estimators with poor finite sample behavior and whose asymptotic distribution, if it exists, may be non-Gaussian. See \citet{Ma-Wang_2020_JASA} for an example in the context of inverse-probability weighting estimators, which includes randomized experiments as a special case.

Next, let
\[\hat{\sigma}^2(\mathbf{a})=\frac{\sum_{g=1}^G\sum_{i=1}^{n+1}(Y_{ig}-\hat{\mu}(\mathbf{a}))^2\1_{ig}(\mathbf{a})}{N(\mathbf{a})}\1(N(\mathbf{a})>0)\]
be the variance estimator for each $\mathbf{a}$. Then we have the following result.
\begin{theorem}[Consistency and asymptotic normality]\label{thm:inference}
Suppose Assumptions \ref{assu:indep}, \ref{assu:sampling} and \ref{assu:moments} hold. Under Condition \eqref{eq:cond_cells} from Lemma \ref{lemma:sampsi} and if $|\mathcal{A}_n|=O(G(n+1)\underline{\pi}_n)$, as $G\to\infty$,
\begin{align}\label{eq:consistency}
\begin{split}
\max_{\mathbf{a}\in\mathcal{A}_n}\left\vert\hat{\mu}(\mathbf{a})-\mu(\mathbf{a})\right\vert &=O_\P\left(\sqrt{\frac{\log|\mathcal{A}_n|}{G(n+1)\underline{\pi}_n}}\right),\\
\max_{\mathbf{a}\in\mathcal{A}_n}\left\vert \hat{\sigma}^2(\mathbf{a})-\sigma^2(\mathbf{a})\right\vert&=O_\P\left(\sqrt{\frac{\log|\mathcal{A}_n|}{G(n+1)\underline{\pi}_n}}\right),
\end{split}
\end{align}
and
\begin{align}\label{eq:asynorm}
\max_{\mathbf{a}\in\mathcal{A}_n}\, \sup_{x\in \mathbbm{R}}\left\vert \P\left[\frac{\hat{\mu}(\mathbf{a})-\mu(\mathbf{a})}{\sqrt{\V[\hat{\mu}(\mathbf{a})|\mathbf{A}]}}\le x \right]-\Phi(x)\right\vert = O\left(\frac{1}{\sqrt{G(n+1)\underline{\pi}_n}}\right)
\end{align}
where $\Phi(x)$ is the cdf of a standard normal random variable. 
\end{theorem}
Equation \eqref{eq:consistency} shows that both the average outcome and variance estimators converge in probability to their true values, uniformly over treatment assignments, at the rate $\sqrt{\log|\mathcal{A}_n|/(G(n+1)\underline{\pi}_n)}$. The denominator in this rate can be seen as the minimum expected cell size, whereas the numerator is a penalty for having an increasing number of parameters. 
Equation \eqref{eq:asynorm} bounds the difference between the distributions of the standardized average outcomes estimators and the standard normal distribution, uniformly over the treatment assignments. Under condition \eqref{eq:cond_cells}, $G(n+1)\underline{\pi}_n\to \infty$, which gives asymptotic normality. Furthermore, this bound also reveals the rate at which the distribution of the standardized estimator approaches the standard normal, namely, $\sqrt{G(n+1)\underline{\pi}_n}$. 

Another implication of Theorem \ref{thm:inference} is that the estimators $\hat{\mu}(\mathbf{a})$ have different convergence rates. More precisely, Lemma \ref{lemma:cons_pi} in the supplemental appendix and Markov's inequality imply that, for each $\mathbf{a}$,  $|\hat{\mu}(\mathbf{a})-\mu(\mathbf{a})|=O_\P\left((G(n+1)\pi(\mathbf{a}))^{-1/2}\right)$ and thus the average outcomes for assignments whose probability decreases faster are estimated at a slower rate.

Finally, the condition that $|\mathcal{A}_n|=O(G(n+1)\underline{\pi}_n)$ requires that the number of parameters do not grow faster than the expected sample size in the smaller cell. Notice that when the number of parameters grows linearly, as in an exchangeable treatment rule, this condition is implied by Condition \eqref{eq:cond_cells}. Section \ref{sec:design} in the supplemental appendix provides sufficient conditions for these requirements under two different assignment mechanisms.

\begin{remark}[Inference with many small groups]\label{rmk:smallgroups}
\upshape When the number of units per group is small compared to the total sample size, the effect of group size disappears asymptotically and inference can be based on a large $G$ small $n$ approximation. In this context, $n$, $|\mathcal{A}_n|$ and $\underline{\pi}_n$ are fixed so condition \eqref{eq:cond_cells} holds automatically as long as the number of groups goes to infinity. Consistency and asymptotic normality of the estimators can be achieved under the usual regularity conditions as $G\to\infty$, and the variance estimator can easily account for both heteroskedasticity and intragroup correlation using standard techniques. The empirical application in Section \ref{sec:empapp} fits into this scenario. $\square$
\end{remark}

\begin{remark}[Inference under misspecified treatment rules]
\upshape Theorem \ref{thm:inference} relies on the true treatment rule $h_0(\cdot)$ being coarser than $h(\cdot)$, which rules out misspecification of $h(\cdot)$. As discussed in Section \ref{sec:identification}, in some cases the researcher may want to posit a coarse treatment rule $h(\cdot)$ to use as a summary measure of direct and spillover effects. These cases are not considered in Theorem \ref{thm:inference}. In such cases, however, as long as $h(\cdot)$ is chosen such that $|\mathcal{A}_n|$ does not increase with the sample size (and therefore the probabilities $\pi(\mathbf{a})$ are fixed), the problem reduces to estimating a finite vector of means. Thus, consistency and asymptotic normality for the vector of parameters of interest follows from existing methods as $G\to\infty$ and $n\to\infty$ (or $n$ fixed) and allowing for within-group dependence \citep[see e.g.][]{Hansen-Lee_2018_JoE}. This applies to cases such as pooled parameters (Remark \ref{rmk:pooled}) and partial population experiments (Remark \ref{rmk:ppexp}). $\square$
\end{remark}

\begin{remark}[Connection to multi-valued treatments]
\upshape This asymptotic framework can also be applied to multi-valued treatments setting \citep{Imbens_2000_BKA} where $\mathbf{A}_{ig}$ corresponds to the treatment of unit $i$ in group $g$. Estimation and inference for multi-valued treatments taking on a finite number of values have been extensively analyzed in a variety of contexts \citep[see e.g.][and references therein]{Cattaneo_2010_JoE,Farrell_2015_JoE,Ao-etal_2020_JBES}. The results in this section complement this literature in two ways. First, I consider double-array asymptotics in which the number of treatment values is allowed to grow with the sample size. Second, the results account for the dependence between treatment assignments $\mathbf{A}_{ig}$ and $\mathbf{A}_{jg}$ of units in the same group, a feature that is specific to the spillovers setting. $\square$
\end{remark}


\subsection{Bootstrap approximation}

An alternative approach to perform inference in this setting is the bootstrap. Since the challenge for inference is that cells can have too few observations for the normal distribution to provide a good approximation, the wild bootstrap \citep{Shao_book} can offer a more accurate approximation when groups are relatively large. One way to implement this type of bootstrap is to define weights $w_{ig}\in\{-1,1\}$ with probability $1/2$ independently of the sample. The bootstrap estimator for $\mu(\mathbf{a})$ is given by:
\[\hat{\mu}^*(\mathbf{a})=\frac{\sum_{g=1}^G\sum_{i=1}^{n+1} Y^*_{ig}\1_{ig}(\mathbf{a})}{N(\mathbf{a})}\]
whenever the denominator is non-zero, where 
\[Y^*_{ig}\1_{ig}(\mathbf{a})=(\bar{Y}(\mathbf{a})+(Y_{ig}-\bar{Y}(\mathbf{a}))w_{ig})\1_{ig}(\mathbf{a})=(\bar{Y}(\mathbf{a})+\hat{\varepsilon}_{ig}w_{ig})\1_{ig}(\mathbf{a})\]
In what follows, $\P^*[\cdot]$ denotes a probability calculated over the distribution of $w_{ig}$, conditional on the sample, and $\E^*[\cdot]$ and $\V^*[\cdot]$ the expectation and variance calculated over $\P^*[\cdot]$. The validity of the wild bootstrap is established in the following theorem.

\begin{theorem}[Wild bootstrap]\label{thm:bootstrap}
Under Assumptions \ref{assu:indep}, \ref{assu:sampling} and \ref{assu:moments},
\[\max_{\mathbf{a}\in\mathcal{A}_n}\sup_{x\in\mathbbm{R}}\left\vert \P^*\left[\frac{\hat{\mu}^*(\mathbf{a})-\hat{\mu}(\mathbf{a})}{\sqrt{\V^*[\hat{\mu}^*(\mathbf{a})}]}\le x\right]-\P\left[\frac{\hat{\mu}(\mathbf{a})-\mu(\mathbf{a})}{\sqrt{\V[\hat{\mu}(\mathbf{a})|\mathbf{A}]}}\le x\right]\right\vert\to_\P 0.\]
\end{theorem}

This theorem shows that the wild bootstrap can be used to approximate the distribution of the estimators as an alternative to the standard normal, which may not be accurate when cells have few observations. The performance of the wild bootstrap is illustrated in Section \ref{sec:simulation} using simulation data.


\section{Simulations}\label{sec:simulation}

This section illustrates the above results in a simulation setting. The outcome is binary and exchangeable, and generated by the following DGP:
\[\P[Y_{ig}(d,s)=1]=\mu(d,s)=0.75+0.13 \times d + 0.12 \times(1-d)\1(s>0)\]
where the spillover effect on an untreated unit is equal to $0.12$ whenever at least one neighbor is treated, and zero for treated units. This DGP is based on the empirical application in Section \ref{sec:empapp}.

The simulations consider two assignment mechanisms. First, \textit{simple random assignment} (SR), where treatment is assigned independently with $\P[D_{ig}=1]=0.5$. Second, \textit{two-stage randomization with fixed margins} (2SR-FM), where in the first stage groups are assigned the total number of treated units between $t=0,1,2,\ldots,n+1$ with probability $q_t$, and then treated units are chosen randomly within each group. The probabilities $q_t$ are chosen so that sample sizes across all assignments are similar; see Section \ref{app:2sr} in the supplemental appendix for further details. Corollaries \ref{coro:simplerand} and \ref{coro:2SR-FM} in the supplemental appendix show that when peers are exchangeable, under SR, condition \eqref{eq:cond_cells} and the conditions in Theorem \ref{thm:inference} hold whenever $(n+1)/\log G\to 0$, whereas under 2SR-FM the conditions hold when $\log (n+1)/\log G\to 0$. Because the second condition is weaker, the normal approximation is expected to perform better for 2SR-FM as groups get larger.

The parameter of interest will be $\theta_0(n)=\E[Y_{ig}(0,n)]-\E[Y_{ig}(0,0)]$, which is the average spillover effect for an untreated unit with all peers treated. In this simulation, $\theta_0(n)=0.12$. This parameters can be seen as a ``worst-case scenario'' given that the probability of the assignment $(D_{ig},S_{ig})=(0,n)$ is one of the smallest (in fact, the smallest under 2SR-FM). The spillover effect estimator is the difference in cell means:
\[\hat{\theta}_0(n)=\frac{\sum_{g=1}^{G}\sum_{i=1}^{n+1} Y_{ig}\1_{ig}(0,n)}{N(0,n)}-\frac{\sum_{g=1}^{G}\sum_{i=1}^{n+1} Y_{ig}\1_{ig}(0,0)}{N(0,0)}\]
whenever $N(0,n)>1$ and $N(0,0)>1$, so that both the estimator and its standard error can be calculated. When at least one of the cells has one or zero observations, the estimator is undefined. 

\begin{figure}
\begin{center}\caption{Coverage rate of the 95\% confidence interval.\label{fig:coverage}}
  \begin{subfigure}[b]{0.45\textwidth}
  \includegraphics[width=\textwidth]{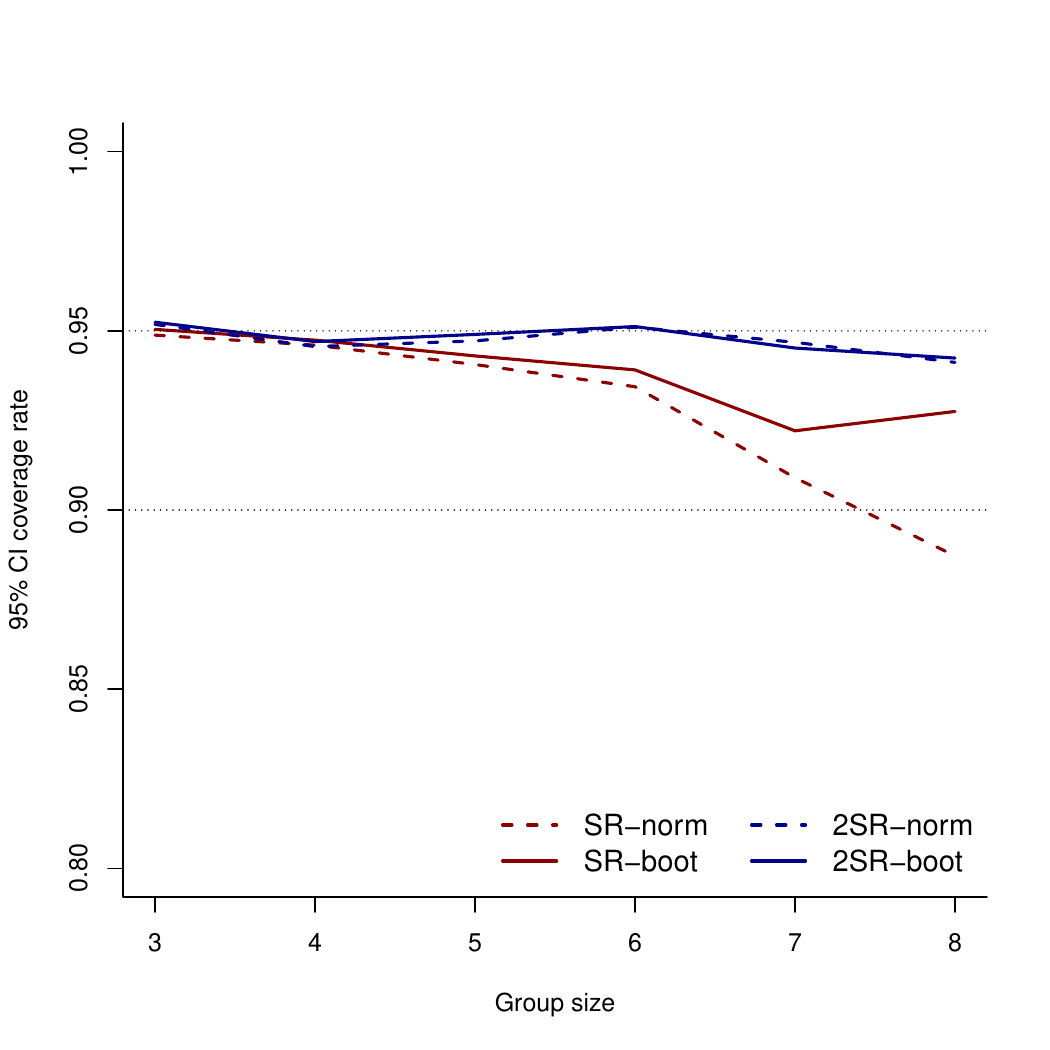}
  \caption{$\hat{\theta}_0(n)$, $G=300$}
  \end{subfigure}
  \begin{subfigure}[b]{0.45\textwidth}
  \includegraphics[width=\textwidth]{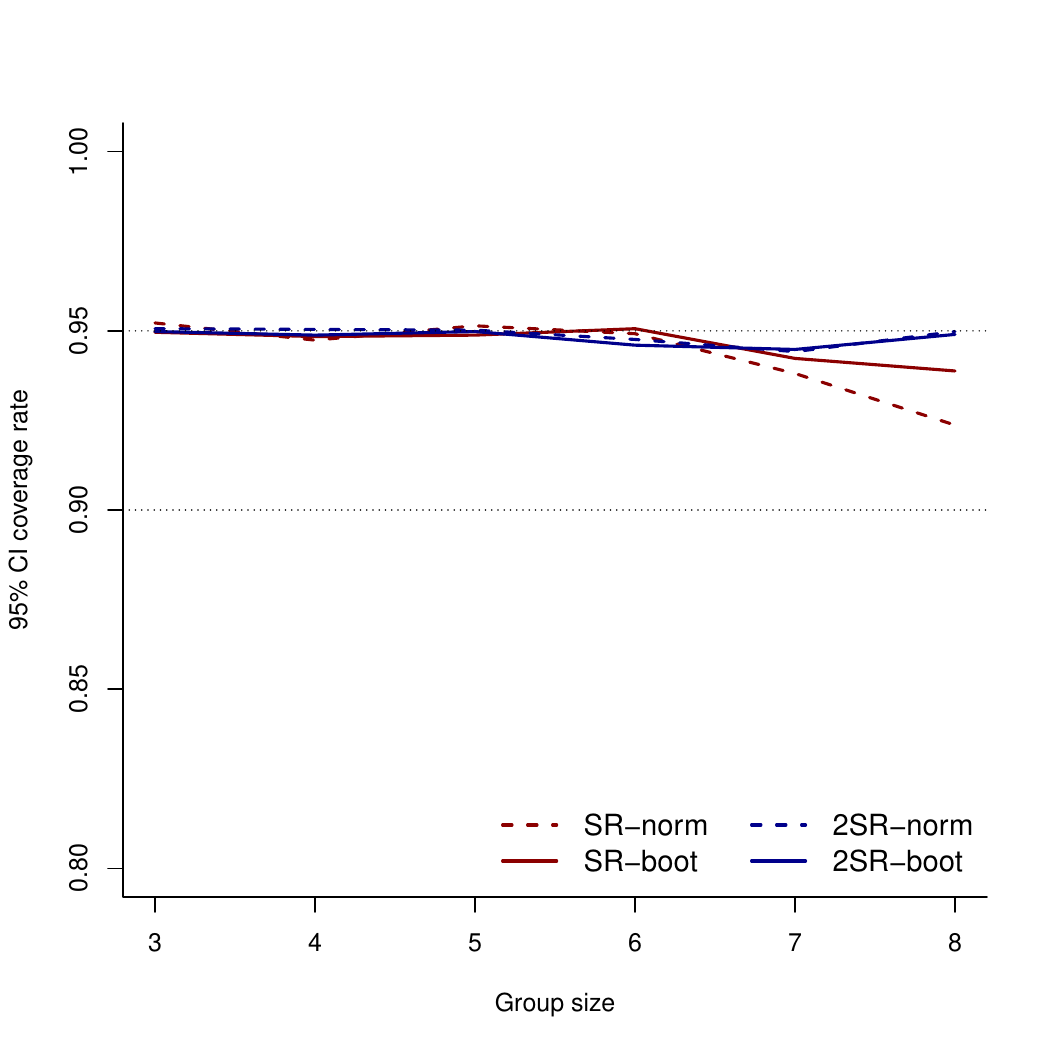}
  \caption{$\hat{\theta}_0(n)$, $G=600$}
  \end{subfigure}
\end{center}
\vspace{-0.05in}\footnotesize\textbf{Notes}: the dashed lines show the coverage rate of the 95\% confidence interval for $\theta_n(0)$ based on the normal approximation under simple random assignment (red line) and two-stage randomization (blue line) for a sample with 300 (left) and 600 (right) groups. The solid lines show the coverage rates for the confidence interval constructed using wild bootstrap. 
\end{figure}

Table \ref{tab:simul_results_300} presents the results for a sample with 300 groups, for group sizes $n+1\in\{3,4,\ldots,8\}$. The upper panel shows the results under SR while the lower panel corresponds to the 2SR-FM assignment. In each panel, the first row gives the value of the condition to achieve consistency and asymptotic normality; intuitively, the closer this value is to zero, the better the normal approximation should be. The second and third rows show the bias and the variance of $\hat{\theta}_0(n)$, calculated over the values of the simulated estimates conditional on the estimate being well defined (i.e. when the cells have enough observations to calculate the estimator and its variance). Rows four to seven show the coverage rate and average length of a 95\% confidence intervals based on the normal approximation and the wild bootstrap.  The eighth row gives the proportion of the simulations in which the estimator or its standard error could not be calculated due to insufficient number of observations. Finally the last two rows show the average sample size in the two assignment cells of interest. Coverage rates are also shown in Figure \ref{fig:coverage}.

\begin{table}\renewcommand{\arraystretch}{1.2}
\begin{center}\caption{Simulation results, $G=300$}\label{tab:simul_results_300}
\input{simul_results_300.txt}
\end{center}
\vspace{-0.05in}\footnotesize\textbf{Notes}: simulation results for $G=300$ groups. The second and third rows in each panel show the bias and variance of $\hat{\theta}_0(n)$. The fourth to seventh rows show the coverage rate and average length of a normal-based and wild-bootstrap-based $95\%$ confidence intervals, respectively. The eighth row shows the proportion of simulations in which $\hat{\theta}_0(n)$ is undefined due to the small number of observations in the corresponding cell. The ninth and tenth rows show the average sample size in the two assignment cells of interest. Results from 5,000 simulations with 1,000 bootstrap replications.
\end{table}

\begin{table}\renewcommand{\arraystretch}{1.2}
\begin{center}\caption{Simulation results, $G=600$}\label{tab:simul_results_600}
\input{simul_results_600.txt}
\end{center}
\vspace{-0.05in}\footnotesize\textbf{Notes}: simulation results for $G=600$ groups. The second and third rows in each panel show the bias and variance of $\hat{\theta}_0(n)$. The fourth to seventh rows show the coverage rate and average length of a normal-based and wild-bootstrap-based $95\%$ confidence intervals, respectively. The eighth row shows the proportion of simulations in which $\hat{\theta}_0(n)$ is undefined due to the small number of observations in the corresponding cell. The ninth and tenth rows show the average sample size in the two assignment cells of interest. Results from 5,000 simulations with 1,000 bootstrap replications.
\end{table}

The simulations reveal that under simple randomization, the estimator performs well for $n\le 4$, with no bias and coverage rates very close to 95\% for both the normal approximation and the wild bootstrap. When $n>4$, however, coverage rates decrease in both cases, more rapidly so for the normal confidence interval whose coverage rate drops to 88\% for $n=7$. While the coverage of the bootstrap confidence interval also decreases, it remains closer to 95\% compared to the normal approximation. The table also shows that under this assignment mechanism, the corresponding sample sizes decrease very rapidly in the relevant assignment cells. When $n=7$, each mean is calculated using 9 observations on average, and is undefined in about 32\% of the simulations. 

On the other hand, under two-stage randomization, both the normal and the wild bootstrap confidence intervals perform equally well and coverage remains very close to 95\%. As shown in the last two rows, two-stage randomization ensures much larger sample sizes in the assignment cells of interest compared to simple randomization.

Table \ref{tab:simul_results_600} shows the same results for a sample with 600 groups. As expected, the estimator and confidence intervals show better performance compare to the case with $G=300$.

\section{Empirical illustration}\label{sec:empapp}

In this section I reanalyze the data from \citet{Barrera-Osorio-etal_2011_AEJ}. The authors conducted a pilot experiment designed to evaluate the effect of a conditional cash transfer program, \textit{Subsidios Condicionados a la Asistencia Escolar}, in Bogot{\'a}, Colombia. The program aimed at increasing student retention and reducing drop-out and child labor. Eligible registrants ranging from grade 6-11 were randomly assigned to treatment and control.\footnote{The experiment had two different treatments that varied the timing of the payments, but, following the authors, I pool the two treatment arms to increase the sample size. See \citet{Barrera-Osorio-etal_2011_AEJ} for details.} The assignment was performed at the student level. In addition to administrative and enrollment data, the authors collected baseline and follow-up data from students in the largest 68 of the 251 schools. This survey contains attendance data and was conducted in the household. As shown in Table \ref{tab:BO_hhsize}, 1,594 households have more than one registered child (rows labeled 2 to 5), and since the treatment was assigned at the child level, this gives variation in the number of treated children per household. This can be seen in Table \ref{tab:BO_ntreat}, which shows that the number of treated children varies from 0 to 5.

I analyze direct and spillover effects restricting the sample to households with three registered siblings, which gives a total of 168 households and 504 observations. The outcome of interest is school attendance. Because groups are very small in this case, inference can be conducted using standard methods (see Remark \ref{rmk:smallgroups}).

\begin{table}
\begin{minipage}{.5\textwidth}
\begin{center}\caption{Distribution of household size}\label{tab:BO_hhsize}
\input{BO_hhsize.txt}
\end{center}
\end{minipage}
\begin{minipage}{.5\textwidth}
\begin{center}\caption{Treated per household}\label{tab:BO_ntreat}
\input{BO_ntreat.txt}
\end{center}
\end{minipage}
\footnotesize\textbf{Notes}: Table \ref{tab:BO_hhsize} indicates the frequencies of household size (i.e. the number of eligible children per household) and Table \ref{tab:BO_ntreat} indicates the frequencies of the number of treated children per household in the sample collected by \citet{Barrera-Osorio-etal_2011_AEJ}.
\end{table}

I start by estimating the average direct and spillover effects exploiting variation in the number of treated siblings using the following regression:
\begin{equation}\label{eq:reg}
\E[Y_{ig} | D_{ig},S_{ig}] = \alpha+ \tau D_{ig}+\sum_{s=1}^{n_g} \theta_0(s)\1(S_{ig}=s)(1-D_{ig})+\sum_{s=1}^{n_g} \theta_1(s)\1(S_{ig}=s)D_{ig}
\end{equation}
Because this regression is saturated, it follows that:
\[\tau=\E[Y_{ig} | D_{ig}=1,S_{ig}=0]-\E[Y_{ig} | D_{ig}=0,S_{ig}=0]\]
and
\[\theta_d(s)=\E[Y_{ig} | D_{ig}=d,S_{ig}=s]-\E[Y_{ig} | D_{ig}=d,S_{ig}=s].\]
Lemma \ref{lemma:identif} provides two alternative ways to interpret these estimands. If siblings are exchangeable, so that average potential outcomes take the form $\E[Y_{ig}(d,s)]$, then $\tau=\E[Y_{ig}(1,0)-Y_{ig}(0,0)]$ is the average direct effect of the treatment on a child with no treated siblings, whereas $\theta_d(s)=\E[Y_{ig}(d,s)-Y_{ig}(d,0)]$ is the average spillover effect of having $s$ treated siblings under own treatment status $d$. In this application, exchangeability may be a reasonable assumption if parents make schooling decisions based on how many of their children are treated (for example, to determine whether the cash transfer covers the direct and opportunity costs of sending their children to school), regardless of which of their children are treated.

On the other hand, if exchangeability does not hold, Lemma \ref{lemma:identif} shows that the parameters from Equation \eqref{eq:reg} combine weighted averages of average potential outcomes. For example, if siblings have a certain (possibly unknown) ordering so that the true average potential outcomes take the form $\E[Y_{ig}(d,\mathbf{d}_g)]$, then:
\[\tau=\E[Y_{ig}(1,\mathbf{0}_g)-Y_{ig}(0,\mathbf{0}_g)]\]
and 
\[\theta_d(s)=\sum_{\mathbf{d}_g:\mathbf{1}_g'\mathbf{d}_g=s}\E[Y_{ig}(d,\mathbf{d}_g)-Y_{ig}(d,\mathbf{0}_g)]\P[\mathbf{D}_{(i)g}=\mathbf{d}_g|D_{ig}=d,S_{ig}=s]\]
which averages the spillover effects over the values of $\mathbf{D}_{(i)g}$ that are consistent with $S_{ig}=s$. Thus, Equation \eqref{eq:reg} provides a way to summarize the direct and spillover effects of the program even when the true structure of the average potential outcomes is unknown. In addition, Tables \ref{tab:BO_results_nonexchange_age} and \ref{tab:BO_results_nonexchange_gender} provide additional results that explore this issue further and provide a way to test exchangeability over different dimensions.

The estimates from Equation \eqref{eq:reg} are shown in the third panel, ``\textbf{Full}'', of Table \ref{tab:BO_results}. These estimates suggest a positive direct effect of the treatment of 16.4 percentage points, significant at the 5 percent level, with almost equally large spillover effects on the untreated units. More precisely, the estimated effect on an untreated kid of having one treated sibling is 14.6 percentage points, while the effect of having two treated siblings is 14 percentage points. The hypothesis that $\theta_0(1)=\theta_0(2)$ cannot be rejected, which suggests some form of crowding-out: given that one sibling is treated, treating one more sibling does not affect attendance. These facts are consistent with the idea that, for example, the conditional cash transfer alleviates some financial constraint that was preventing the parents from sending their children to school regularly, or with the program increasing awareness on the importance of school attendance, since in these cases the effect occurs as soon as one kid in the household is treated, and does not amplify with more treated children. 

On the other hand, spillover effects on treated children are much smaller in magnitude and negative. The fact that these estimates are negative does not mean that the program hurts treated children, but that treating more siblings reduces the benefits of the program. For example, the effect of being treated with two treated siblings, compared to nobody treated, can be estimated by $\hat{\tau}+\hat{\theta}_1(2)\approx 0.113$. Thus, a treated kid with two treated siblings increases her attendance in 11 percentage points starting from a baseline in which nobody in the household is treated. 

In all, the estimates suggest large and positive direct and spillover effects on the untreated, with some evidence of crowding-out between treated siblings.\footnote{These empirical findings differ from those in \citet{Barrera-Osorio-etal_2011_AEJ}, who find evidence of negative spillover effects. Their results are calculated over a different sample, since the authors focus on households with two registered children, whereas I consider households with three registered children. Differences in estimated direct and spillover effects may be due to heterogeneous effects across household sizes. For instance, in this sample, households with more registered children have lower income on average, so they may benefit differently from the program.}

\afterpage{%
\clearpage
\begin{landscape}
\begin{table}[h]
\begin{center}\caption{Estimation results}\label{tab:BO_results}
\resizebox{\textwidth}{!}{\input{BO_regs_comparison.txt}}

\vspace{0.1cm}\scriptsize\textbf{Notes}: S.e. clustered at the household level. Regressions include school FE. ***$p<0.01$,**$p<0.05$,*$p<0.1$.
\end{center}
\end{table}
\end{landscape}
\clearpage
}

\subsection{Difference in means}

Using the results from the nonparametric specification as a benchmark, I now estimate the effects of the program using the  difference in means analyzed in Section \ref{sec:dm}. The left panel, ``\textbf{Diff. Means}'', of Table \ref{tab:BO_results} shows the difference in means, calculated as the OLS estimator for $\beta_{\mathsf{D}}$ in Equation \eqref{eq:diffmeans}. The results show that the difference in means is practically zero and not significant. Hence, by ignoring the presence of spillover effects, a researcher estimating the effect of the program in this way would conclude that the treatment has no effect. 

This finding is due to the fact that the difference in means combines all the effects in the third panel into a single number, as shown in Theorem \ref{thm:dm}. From Table \ref{tab:BO_results}, the estimated spillover effects in this case are larger under control that under treatment, and have different signs, so $\hat{\theta}_1(s)-\hat{\theta}_0(s)<0$. Therefore, the spillover effects push the difference in means towards zero in this case.

\subsection{Reduced-form linear-in-means regression}

Next, I estimate the effects using the RF-LIM regression analyzed in Section \ref{sec:lim}. The estimates from Equation \eqref{eq:lim} are given in the first column of the middle panel, ``\textbf{Linear-in-Means}'', in Table \ref{tab:BO_results}. The estimates reveal very small and statistically insignificant direct and spillover effects, substantially different from the results using Equation \eqref{eq:reg}. 

Theorem \ref{thm:lim} and Corollary \ref{coro:lim} show that a RF-LIM regression implicitly imposes linearity of spillover effects and may suffer from misspecification when spillover effects are nonlinear. The estimates from the full nonparametric specification show that spillover effects are highly nonlinear in this case, which explains why the RF-LIM regression fails to recover these effects. 

The second column in the RF-LIM panel presents the estimates from the interacted RF-LIM regression shown in Equation \eqref{eq:lim_inter}. The results reveal that separately estimating the spillover effects for treated and controls mitigates the misspecification in this case, and the estimates are closer to the ones from the nonparametric specification, although the fact that 0.169 is not a weighted average of 0.146 and 0.14 suggests that some extrapolation bias remains due to the nonlinearity of spillover effects. 

\subsection{Pooled effects}

I now illustrate how to estimate pooled effects by averaging over the possible number of treated siblings (2 and 3 in this case). For this, I estimate the following regression:
\[Y_{ig}=\alpha_p+\tau D_{ig}+\gamma_p^0\1(S_{ig}>0)(1-D_{ig})+\gamma_p^1\1(S_{ig}>0)D_{ig}+\nu_{ig}\]
where
\[\tau=\E[Y_{ig}|D_{ig}=1,S_{ig}=0]-\E[Y_{ig}|D_{ig}=0,S_{ig}=0]\]
and
\[\gamma_p^d=\sum_{s=1}^2\theta_d(s)\P[S_{ig}=s|D_{ig}=d,S_{ig}>0]\]
where $\theta_d(s)$ is defined in Equation \eqref{eq:reg} (see also Remark \ref{rmk:pooled}). From Table \ref{tab:BO_results} we can see that the estimated pooled spillover effects are $0.144$ for controls and $-0.045$ for treated, which, as expected, lie between the effects found with the saturated regression. These results illustrate how this type of pooling can provide a useful summary of spillover effects, which may be a feasible alternative when the total number of spillover effects is too large or cell sizes are small to estimate them separately.

\subsection{Non-exchangeable peers}

Next, I illustrate how to relax the exchangeability assumption in two ways. First, I define an ordering between siblings by looking at differences (in absolute value) in ages, defining sibling 1 as the sibling closest in age and sibling 2 as the sibling farthest in age. Then, estimation is conducted by simply adding indicator variables for the possible different assignments. Table \ref{tab:BO_results_nonexchange_age} shows the estimates from this specification. The estimates reveal similar results to Table \ref{tab:BO_results}, with a direct effect of 0.165, spillover effects on the untreated ranging from 0.133 to 0.14 and spillover effects on the treated ranging from -0.039 to -0.051.

In fact, exchangeability can be tested by assessing whether the spillover effects of siblings 1 and 2 are the same, which in this case amounts to testing equality of coefficients between rows 2 and 3, and between rows 5 and 6. The test statistic and corresponding p-value from this test are given in the last two rows of the table, where it is clear that exchangeability cannot be rejected in this case, although this could be due to low statistical power given the relatively small sample size. 

\begin{table}[h]
\begin{center}\caption{Estimation results by age}\label{tab:BO_results_nonexchange_age}
\resizebox{.5\textwidth}{!}{\input{BO_regs_nonexchange_age.txt}}

\vspace{0.1cm}\footnotesize\textbf{Notes}: Cluster-robust s.e. Regressions include school FE. ***$p<0.01$,**$p<0.05$,*$p<0.1$.
\end{center}
\end{table}

Finally, I consider the case in which the effect of treated siblings depends on whether siblings are male or female, allowing for non-exchangeable siblings based on gender. The results are shown in Table \ref{tab:BO_results_nonexchange_gender}. In this table, $S_{ig}^m$ denotes the number of male treated siblings and $S_{ig}^f$ denotes the number of female treated siblings. The results are qualitatively similar, with some suggestive evidence of slightly larger spillover effects from female siblings. The hypothesis that the coefficients are the same cannot be rejected, although again the sample may be too small to draw precise conclusions about sibling exchangeability. 

\begin{table}[h]
\begin{center}\caption{Estimation results by gender}\label{tab:BO_results_nonexchange_gender}
\resizebox{.5\textwidth}{!}{\input{BO_regs_nonexchange_gender.txt}}

\vspace{0.1cm}\footnotesize\textbf{Notes}: Cluster-robust s.e. Regressions include school FE. ***$p<0.01$,**$p<0.05$,*$p<0.1$.
\end{center}
\end{table}


\section{Discussion}\label{sec:conclusion}

The findings in this paper offer several takeaways for analyzing spillover effects in randomized experiments. First, commonly-analyzed estimands such as the difference in means and RF-LIM coefficients implicitly impose strong assumptions on the structure of spillover effects, and are therefore not recommended as they generally do not have a causal interpretation. On the other hand, the full vector of spillover effects is identifiable whenever the experimental design generates enough variation in the number of treated units in each group and the researcher assumes a treatment rule that is flexible enough.

Second, while nonparametric estimation of all direct and spillover effects can give a complete picture of the effects of the treatment, it can be difficult to implement in practice when groups are large. As a guideline to determine in which cases spillover effects can be estimated nonparametrically, Theorem \ref{thm:inference} formalizes the notion of a ``sufficiently large sample'' in this context, and provides a way to assess the performance of the different types of treatment effect estimators depending on the number of groups, number of parameters of interest and treatment assignment mechanism. As an alternative, pooled estimands can recover weighted averages of spillover effects with known weights for which inference can be conducted under standard conditions.

The supplemental appendix discusses several important issues that can be further developed in future research. Sections \ref{app:group_size} and \ref{app:covariates} discuss extensions to unequal group sizes and the inclusion of covariates. The results in Section \ref{sec:estimation} and the simulations in Section \ref{sec:simulation} highlight the fact that the rate of convergence of the proposed estimators depend on the experimental design. This suggests that these results can be used to rank treatment assignment mechanisms, and this has implications for experimental design, as discussed in Section \ref{sec:design}. 

The analysis in this paper leaves several open questions to be explored. One example is allowing for endogenous group formation. The identification results in this paper follow through when groups are endogenously formed, as long as they are formed before the treatment is assigned and their structure is not changed by the treatment (inference may require further assumptions to account for possible overlap between groups). On the other hand, when the structure of the group is affected by the treatment, the treatment can affect outcomes through direct effects, through spillover effects given the network, and through changing the network structure. In such cases, while it is possible to identify the ``overall'' effect \citep{Kline-Tamer_2019}, random assignment of the treatment is generally not enough to separately point identify these different effects, and further assumptions are needed. On the other hand, the findings in this paper can be generalized to settings where the researcher does not have precise control on treatment take-up. In this direction, \citet{Vazquez-Bare_2020_IV} analyzes instrumental variable methods that can be applied to RCTs with imperfect compliance, where treatment receipt is endogenous, or more generally in quasi-experimental settings.


\newpage
\bibliographystyle{econometrica}
\bibliography{Spillovers_references}


\newpage

\begin{appendices}

\section{Endogenous Effects and Structural Models}\label{app:structural}

Consider the structural model:
\[Y_{ig}=\phi(D_{ig},\mathbf{D}_{(i)g})+\gamma \bar{Y}_{g}^{(i)}+u_{ig}.\]
which assumes additive separability between the functions that depend on treatment assignments and on outcomes. In this model, $\phi(1,\mathbf{d}_g)-\phi(0,\mathbf{d}_g)$, measures the direct effect of the treatment, $\phi(d,\mathbf{d}_g)-\phi(d,\mathbf{\tilde{d}}_g)$ measures the spillover effects of peers' treatments, commonly known as exogenous or contextual effects, and $\gamma$ measures the endogenous effect.

Suppose that the assumptions from Corollary \ref{coro:lim} hold, so that the treatment vector is randomly assigned, peers are exchangeable and spillover effects are linear. By exchangeability,
\begin{align*}
\phi(D_{ig},\mathbf{D}_{(i)g})=\phi(D_{ig},S_{ig})&=\sum_{s=0}^{n_g}\tilde{\beta}_s \1(S_{ig}=s)(1-D_{ig})+\sum_{s=0}^{n_g}\tilde{\delta}_s \1(S_{ig}=s)D_{ig}\\
&=\tilde{\beta}_0 + (\tilde{\delta}_0-\tilde{\beta}_0)D_{ig}+\sum_{s=1}^{n_g}(\tilde{\beta}_s-\tilde{\beta}_0) \1(S_{ig}=s)(1-D_{ig})\\
&+\sum_{s=1}^{n_g}(\tilde{\delta}_s-\tilde{\delta}_0) \1(S_{ig}=s)D_{ig}
\end{align*}
where the second equality is without loss of generality because all the variables are discrete, and where $\tilde{\beta}_s=\phi(0,s),\quad \tilde{\delta}_s=\phi(1,s)$.
Let $\alpha=\tilde{\beta}_0$, $\beta=\tilde{\delta}_0-\tilde{\beta}_0$, $\gamma_s^0=\tilde{\beta}_s-\tilde{\beta}_0$, $\gamma_s^1=\tilde{\delta}_s-\tilde{\delta}_0$ and rewrite the above model as:
\[\phi(D_{ig},S_{ig})=\alpha + \beta D_{ig}+\sum_{s=1}^{n_g}\gamma_s^0\1(S_{ig}=s)(1-D_{ig})+\sum_{s=1}^{n_g}\gamma_s^1\1(S_{ig}=s)D_{ig}.\]

Next, by linearity of spillover effects, $\gamma_0^d=\kappa_d s$ and $\sum_{s=1}^{n_g}\gamma_s^d\1(S_{ig}=s)=\kappa_d\sum_{s=1}^{n_g}s\1(S_{ig}=s)=\kappa_d S_{ig}$. Therefore,
\[Y_{ig}=\alpha+\beta D_{ig}+\kappa_0S_{ig}(1-D_{ig})+\kappa_1 S_{ig}D_{ig}+\gamma \bar{Y}_{g}^{(i)}+u_{ig}.\]
In addition, suppose that contextual effects are equal between treated and controls so that $\kappa_0=\kappa_1=\kappa$. The model then reduces to:
\begin{align*}
Y_{ig}&=\alpha+\beta D_{ig}+\kappa S_{ig}+\gamma \bar{Y}_{g}^{(i)}+u_{ig}=\alpha+\beta D_{ig}+\kappa n_g \bar{D}_{g}^{(i)}+\gamma \bar{Y}_{g}^{(i)}+u_{ig}.
\end{align*}
Noting that $\kappa$ can be a function of $n_g$, $\kappa=\kappa(n_g)$, let $\theta=\kappa(n_g)n_g$ where the dependence on $n_g$ is left implicit, so that:
\begin{align*}
Y_{ig}&=\alpha+\beta D_{ig}+ \theta\bar{D}_{g}^{(i)}+\gamma \bar{Y}_{g}^{(i)}+u_{ig}
\end{align*}
which is a standard LIM model where $\beta$ is the direct effect of the treatment, $\theta$ is the exogenous or contextual effect and $\gamma$ is the endogenous effect. 

Next, note that $\bar{Y}_{g}^{(i)}=\frac{n_g+1}{n_g}\bar{Y}_g-\frac{Y_{ig}}{n_g}$ which implies that:
\[Y_{ig}\left(1+\frac{\gamma}{n_g}\right)=\alpha+\beta D_{ig}+ \theta\bar{D}_{g}^{(i)}+\gamma \left(\frac{n_g+1}{n_g}\right)\bar{Y}_g+u_{ig}\]
and 
\[\bar{Y}_g=\alpha+\beta \bar{D}_{g}+ \theta\bar{D}_{g}+\gamma \bar{Y}_{g}+\bar{u}_{g}.\]
The last equation implies that, as long as $\gamma\ne 1$,
\begin{align*}
\bar{Y}_g&=\frac{\alpha}{1-\gamma}+\frac{\beta+\theta}{1-\gamma}\bar{D}_g+\frac{\bar{u}_g}{1-\gamma}\\
&=\frac{\alpha}{1-\gamma}+\frac{\beta+\theta}{1-\gamma}\left(\frac{1}{n_g+1}\right)D_{ig}+\frac{\beta+\theta}{1-\gamma}\left(\frac{n_g}{n_g+1}\right)\bar{D}_{g}^{(i)}+\frac{\bar{u}_g}{1-\gamma}
\end{align*}
so plugging back:
\begin{align*}
Y_{ig}\left(1+\frac{\gamma}{n_g}\right)&=\alpha+\gamma \left(\frac{n_g+1}{n_g}\right)\frac{\alpha}{1-\gamma}\\
&+\beta D_{ig} +\gamma \left(\frac{n_g+1}{n_g}\right)\frac{\beta+\theta}{1-\gamma}\left(\frac{1}{n_g+1}\right)D_{ig}\\
&+ \theta\bar{D}_{g}^{(i)}+\gamma \left(\frac{n_g+1}{n_g}\right)\frac{\beta+\theta}{1-\gamma}\left(\frac{n_g}{n_g+1}\right)\bar{D}_{g}^{(i)}\\
&+u_{ig}+\gamma \left(\frac{n_g+1}{n_g}\right)\frac{\bar{u}_g}{1-\gamma}
\end{align*}
After some simplifications,
\begin{align*}
Y_{ig}\left(1+\frac{\gamma}{n_g}\right)&=\left[1+ \left(\frac{n_g+1}{n_g}\right)\frac{\gamma}{1-\gamma}\right]\alpha+\left[\beta +\frac{\gamma}{1-\gamma}\cdot\frac{\beta+\theta}{n_g}\right]D_{ig}\\
&+\left[\theta+\gamma \cdot \frac{\beta+\theta}{1-\gamma}\right]\bar{D}_{g}^{(i)}+u_{ig}+\gamma \left(\frac{n_g+1}{n_g}\right)\frac{\bar{u}_g}{1-\gamma}
\end{align*}
and thus
\[Y_{ig}=\alpha^*+\beta^* D_{ig}+\theta^*\bar{D}_{g}^{(i)}+u^*_{ig} \]
where
\begin{align*}
\alpha^*&=\left[1+ \left(\frac{n_g+1}{n_g}\right)\frac{\gamma}{1-\gamma}\right]\left(1+\frac{\gamma}{n_g}\right)^{-1}\alpha\\
\beta^*&=\left[\beta +\frac{\gamma}{1-\gamma}\cdot\frac{\beta+\theta}{n_g}\right]\left(1+\frac{\gamma}{n_g}\right)^{-1} \\
\theta^*&=\left[\theta+\gamma \cdot \frac{\beta+\theta}{1-\gamma}\right]\left(1+\frac{\gamma}{n_g}\right)^{-1}\\
u^*_{ig}&=u_{ig}\left(1+\frac{\gamma}{n_g}\right)^{-1}+\gamma \left(\frac{n_g+1}{n_g}\right)\left(1+\frac{\gamma}{n_g}\right)^{-1}\frac{\bar{u}_g}{1-\gamma}.
\end{align*}
In this context, random assignment of the treatment implies that $\E[u_{ig}|D_{ig},\mathbf{D}_{(i)g}]=0$ and hence the reduced-form parameters $(\alpha^*,\beta^*,\theta^*)$ are identified. As in any structural LIM model, however, the structural parameters $(\alpha,\beta,\theta,\gamma)$ are not identified without further assumptions.


\section{Assignment Mechanism for 2SR-FM}\label{app:2sr}

In a 2SR-FM assignment mechanism, given a group size $n+1$ groups are assigned to receive $0,1,2,\ldots,n+1$ treated units with probabilities $q_0,q_1,\ldots,q_{n+1}$. Treatment assignments in this case are given by $\mathbf{A}_{ig}=(D_{ig},T_g)$ where $D_{ig}\in\{0,1\}$ and $T_g\in\{0,1,\ldots,n+1\}$, and $\pi(\mathbf{a})=\P[D_{ig}=d|T_g=t]q_t=q_t\left(\frac{t}{n+1}\right)^d\left(1-\frac{t}{n+1}\right)^{1-d}$. When $n+1$ is odd, the choice of $q_t$ is determined by the following system of equations:
\begin{align*}
q_j=q_{n+1-j},\quad j\le \frac{n}{2}\\
q_j=\frac{(n+1)q_0}{j},\quad 0<j\le \frac{n}{2} \\
\sum_jq_j=1.
\end{align*}
The first set of equations imposes symmetry, that is, $\P[T_g=0]=\P[T_g=n+1]$ and so on. The second set of equations makes the expected sample size in the smallest assignment in each group (untreated units in high-intensity treatment groups and vice versa) equal to the expected sample size of pure controls. The solution to this system is given by:
\[q_0\left(1+(n+1)\sum_{j=1}^{\frac{n}{2}} \frac{1}{j}\right)=\frac{1}{2}\]
and the remaining probabilities are obtained from the previous relationships. If $n+1$ is even, the system of equations is given by:
\begin{align*}
q_j=q_{n+1-j},\quad j\le \frac{n-1}{2}\\
q_j=\frac{(n+1)q_0}{j},\quad 0<j\le \frac{n-1}{2} \\
\sum_jq_j=1.
\end{align*}
and the solution is:
\[q_0\left(2+(n+1)\sum_{j=1}^{\frac{n+1}{2}-1} \frac{1}{j}\right)=\frac{1}{2}.\]


\section{Implications for Experimental Design}\label{sec:design}

Theorem \ref{thm:inference} shows that the accuracy of the standard normal to approximate the distribution of the standardized statistic depends on the treatment assignment mechanism through $\underline{\pi}_n$. The intuition behind this result is that the amount of information to estimate each $\mu(\mathbf{a})$ depends on the number of observations facing assignment $\mathbf{a}$, and this number depends on $\pi(\mathbf{a})$. When the goal is to estimate all the $\mu(\mathbf{a})$ simultaneously, the binding factor will be the number of observations in the smallest cell, controlled by $\underline{\pi}_n$. When an assignment sets a value of $\underline{\pi}_n$ that is very close to zero, the normal distribution may provide a poor approximation to the distribution of the estimators. 

When designing an experiment to estimate spillover effects, the researcher can choose distribution of treatment assignments $\pi(\cdot)$. Theorem \ref{thm:inference} provides a way to rank different assignment mechanisms based on their rate of the approximation, which gives a principled way to choose between different assignment mechanisms. 

To illustrate these issues, consider the case of an exchangeable exposure mapping $\mathcal{A}_n=\{(d,s):d=0,1,s=0,1,\ldots,n\}$. The results below compare two treatment assignment mechanisms: simple random assignment (SR) and two-stage randomization with fixed margins (2SR-FM). See Section \ref{app:2sr} for further details on this design.

\begin{corollary}[SR]\label{coro:simplerand}
Under simple random assignment, if:
\begin{equation}\label{eq:simplerand}
\frac{n+1}{\log G}\to 0,
\end{equation}
then $\frac{\log|\mathcal{A}_n|}{G\underline{\pi}_n}\to 0$ and $\frac{|\mathcal{A}_n|}{G(n+1)\underline{\pi}_n}=O(1)$.
\end{corollary}

\begin{corollary}[2SR-FM]\label{coro:2SR-FM}
Under the 2SR-FM mechanism described in Section \ref{sec:design}, if:
\begin{equation}\label{eq:2sr-fm}
\frac{\log(n+1)}{\log G}\to 0,
\end{equation}
then $\frac{\log|\mathcal{A}_n|}{G\underline{\pi}_n}\to 0$ and $\frac{|\mathcal{A}_n|}{G(n+1)\underline{\pi}_n}=O(1)$.
\end{corollary}
In qualitative terms, both results imply that estimation and inference for spillover effects requires group size to be small relative to the total number of groups. Thus, these results formalize the requirement of ``many small groups'' that is commonly invoked, for example, when estimating LIM models. 

Corollary \ref{coro:simplerand} shows that when the treatment is assigned using simple random assignment, group size has to be small relative to $\log G$. Given the concavity of the $\log$ function, this is a strong requirement. Hence, groups have to be very small relative to the sample size for inference to be asymptotically valid. The intuition behind this result is that under a SR, the probability of the tail assignments $(0,0)$ and $(1,n)$ decreases exponentially fast with group size.

On the other hand, Corollary \ref{coro:2SR-FM} shows that a 2SR-FM mechanism reduces the requirement to $\log (n+1)/\log G\approx 0$, so now the log of group size has to be small compared to the log of the number of groups. This condition is much more easily satisfied, which in practical terms implies that a 2SR-FM mechanism can handle larger groups compared to SR. The intuition behind this result is that, by fixing the number of treated units in each group,  a 2SR-FM design has better control on how small the probabilities of each assignment can be, hence facilitating the estimation of the tail assignments. Also note that Condition \eqref{eq:2sr-fm} can be replaced by $n\log n/G\to 0$, $n^2/G=O(1)$.


\section{Unequally-Sized Groups}\label{app:group_size}

To explicitly account for different group sizes, let  $n$ (the total number of peers in each group) take values in $\mathcal{N}=\{n_1,n_2,\ldots,n_K\}$ where $n_k\ge 1$ for all $k$ and $n_1<n_2<\ldots<n_K$. Let the potential outcome be $Y_{ig}(n,d,s(n))$ where $n\in\mathcal{N}$ and $s(n)\in\{0,1,2,\ldots,n\}$. Let $N_g$ be the observed value of $n$, $S_{ig}(n)=\sum_{j\ne i}^nD_{jg}$ and $S_{ig}=\sum_{k=1}^K S_{ig}(n_k)\1(N_g=n_k)$. The independence assumption can be modified to hold conditional on group size:
\[\{Y_{ig}(n,d,s(n)):d=0,1,s(n)=0,1,\ldots,n\}_{i=1}^n\indep \mathbf{D}_g(n))|N_g=n\]
where $\mathbf{D}_g(n)$ is the vector of all treatment assignments when the group size is $n+1$.

Under this assumption, we have that for $n\in\mathcal{N}$ and $s\le n$,
\[\E[Y_{ig}|D_{ig}=d,S_{ig}=s,N_g=n]=\E[Y_{ig}(n,d,s)].\]
The average observed outcome conditional on $N_g=n$ can be written as:
\begin{align*}
\E[Y_{ig}|D_{ig},S_{ig},N_g=n]&=\E[Y_{ig}(n,0,0)]+\tau_0(n)D_{ig}\\
&\quad + \sum_{s=1}^{n}\theta_0(s,n)\1(S_{ig}=s)(1-D_{ig})\\
&\quad + \sum_{s=1}^{n}\theta_1(s,n)\1(S_{ig}=s)D_{ig}
\end{align*}

The easiest approach is to simply run separate analyses for each group size and estimate all the effects separately. In this case, it is possible to test whether spillover effects are different in groups with different sizes. The total number of parameters in this case is given by $\sum_{k=1}^K(n_k+1)$.

In practice, however, there may be cases in which group size has a rich support with only a few groups at each value $n$, so separate analyses may not be feasible. In such a setting, a possible solution is to impose an additivity assumption on group size. According to this assumption, the average direct and spillover effects do not change with group size. For example, the spillover effect of having one treated neighbor is the same in a group with two or three units. Under this assumption,
\begin{align*}
\E[Y_{ig}|D_{ig},S_{ig},N_g]&=\sum_{n\in\mathcal{N}_g} \alpha(n)\1(N_g=n_g) +\tau_0D_{ig}\\
&\quad + \sum_{s=1}^{N_g}\theta_0(s)\1(S_{ig}=s)(1-D_{ig})\\
&\quad + \sum_{s=1}^{N_g}\theta_1(s)\1(S_{ig}=s)D_{ig}
\end{align*}
where the first sum can be seen in practice as adding group-size fixed effects. Then, the identification results and estimation strategies in the paper are valid after controlling for group-size fixed effects. Note that in this case the total number of parameters to estimate is $n_K+K-1$ where $n_K$ is the size of the largest group and $K$ is the total number of different group sizes.

Another possibility is to assume that for any constant $c\in\mathbb{N}$, $Y_{ig}(c\cdot n,d,c\cdot s)=Y_{ig}(n,d,s)$. This assumption allows us to rewrite the potential outcomes as a function of the ratio of treated peers, $Y_{ig}(d,s/n)$. Letting $P_{ig}=S_{ig}/N_g$, all the parameters can be estimated by running a regression including $D_{ig}$, $\1(P_{ig}=p)$ for all possible values of $p>0$ (excluding $p=0$ to avoid perfect collinearity) and interactions. In this case, the total number of parameters can be bounded by $n_1+\sum_{k=2}^K(n_k-1)$. Note that assuming that the potential outcomes depend only on the proportion of treated siblings does not justify including the variable $P_{ig}$ linearly, as commonly done in linear-in-means models.


\section{Including Covariates}\label{app:covariates}

There are several reasons why one may want to include covariates when estimating direct and spillover effects. First, pre-treatment characteristics may help reduce the variability of the estimators and decrease small-sample bias, which is standard practice when analyzing randomly assigned programs. Covariates can also help get valid inference when the assignment mechanisms stratifies on baseline covariates. This can be done by simply augmenting Equation \eqref{eq:reg} with a vector of covariates $\boldsymbol{\gamma}'\mathbf{x}_{ig}$ which can vary at the unit or at the group level. The covariates can also be interacted with the treatment assignment indicators to explore effect heterogeneity across observable characteristics (for example, by separately estimating effects for males and females. 

Second, exogenous covariates can be used to relax the mean-independence assumption in observational studies. More precisely, if $\mathbf{X}_g$ is a matrix of covariates, a conditional mean-independence assumption would be $\E[Y_{ig}(d,\mathbf{d}_g)|\mathbf{X}_g,\mathbf{D}_g]=\E[Y_{ig}(d,\mathbf{d}_g)|\mathbf{X}_g]$ which is a version of the standard unconfoundeness condition. The vector of covariates can include both individual-level and group-level characteristics. 

Third, covariates can be included to make an exposure mapping more likely to be correctly specified. For instance, the exchangeability assumption can be relaxed by assuming it holds after conditioning on covariates, so that for any pair of treatment assignments $\mathbf{d}_g$ and $\mathbf{\tilde{d}}_g$ with the same number of ones, $\E[Y_{ig}(d,\mathbf{d}_g)|\mathbf{X}_g]=\E[Y_{ig}(d,\mathbf{\tilde{d}}_g)|\mathbf{X}_g]$. As an example, exchangeability can be assumed to hold for all siblings with the same age, gender or going to the same school.

All the identification results in the paper can be adapted to hold after conditioning on covariates. In terms of implementation, when the covariates are discrete the parameters of interest can be estimated at each possible value of the matrix $\mathbf{X}_g$, although this strategy can worsen the dimensionality problem. Alternatively, covariates can be included in a regression framework after imposing parametric assumptions, for example, assuming the covariates enter linearly.


\section{Technical Lemmas}

These additional results are used in the proofs of the main results. The proofs of the technical lemmas are given in the supplemental appendix.

\begin{lemma}\label{lemma:cons_pi}
Let $\hat{\pi}(\mathbf{a}):=\sum_g\sum_i \1_{ig}(\mathbf{a})/G(n+1)$. Under the assumptions of Lemma \ref{lemma:sampsi}, for any $\varepsilon>0$,
\[|\mathcal{A}_n|\max_{\mathbf{a}\in\mathcal{A}_n}\P\left[\left\vert \frac{\hat{\pi}(\mathbf{a})}{\pi(\mathbf{a})}-1 \right\vert>\varepsilon \right]\to 0.\] 
\end{lemma}

\begin{lemma}\label{lemma:cons_pi2}
Under the assumptions of Lemma \ref{lemma:cons_pi},
\[\max_{\mathbf{a}\in\mathcal{A}_n}\left\vert \frac{\hat{\pi}(\mathbf{a})}{\pi(\mathbf{a})}-1 \right\vert\to_\P 0.\]
\end{lemma}

\section{Proofs of Main Results}

\paragraph{Proof of Lemma \ref{lemma:identif}}

If $\P[D_{ig}=d,\mathbf{H}_{ig}=\mathbf{h}]>0$,
\begin{align*}
\E[Y_{ig}|D_{ig}=d,\mathbf{H}_{ig}=\mathbf{h}]&=\sum_{\mathbf{h}_0}\E[Y_{ig}|D_{ig}=d,\mathbf{H}_{ig}=\mathbf{h},\mathbf{H}^0_{ig}=\mathbf{h}_0]\\
&\qquad \times \P[\mathbf{H}^0_{ig}=\mathbf{h}_0|D_{ig}=d,\mathbf{H}_{ig}=\mathbf{h}]\\
&=\sum_{\mathbf{h}_0}\E[Y_{ig}(d,\mathbf{h}_0)|D_{ig}=d,\mathbf{H}_{ig}=\mathbf{h},\mathbf{H}^0_{ig}=\mathbf{h}_0]\\
&\qquad \times \P[\mathbf{H}^0_{ig}=\mathbf{h}^0|D_{ig}=d,\mathbf{H}_{ig}=\mathbf{h}]\\
&=\sum_{\mathbf{h}_0}\E[Y_{ig}(d,\mathbf{h}_0)]\P[\mathbf{H}^0_{ig}=\mathbf{h}_0|D_{ig}=d,\mathbf{H}_{ig}=\mathbf{h}]
\end{align*}
where the first equality follows from the law if iterated expectations, the second equality follows by definition of the observed outcomes and the third equality follows from random assignment of the treatment vector given that both $\mathbf{H}^0_{ig}$ and $\mathbf{H}_{ig}$ are deterministic functions of $\mathbf{D}_g$. Finally, if $h_0(\cdot)$ is coarser than $h(\cdot)$, then $\mathbf{H}_{ig}=\mathbf{h}$  uniquely determines the value of $\mathbf{H}^0_{ig}$ and the result follows. $\square$


\paragraph{Proof of Theorem \ref{thm:dm}}

Follows from Lemma \ref{lemma:identif} letting $h(\cdot)$ be a constant function, using the fact that by construction $\beta_\mathsf{D}=\E[Y_{ig}|D_{ig}=1]-\E[Y_{ig}|D_{ig}=0]$. $\square$


\paragraph{Proof of Theorem \ref{thm:lim}}

The coefficients from Equation \eqref{eq:lim} are characterized by the minimization problem:
\[\min_{(\alpha_\ell,\beta_\ell,\gamma_\ell)}\E\left[\left(Y_{ig}-\alpha_\ell-\beta_\ell D_{ig}-\gamma_\ell \bar{D}^{(i)}_g\right)^2\right].\]
The objective function can be rewritten as:
\begin{align*}
\E\left[\left(Y_{ig}-\alpha_\ell-\gamma_\ell \bar{D}^{(i)}_g\right)^2(1-D_{ig})\right]+\E\left[\left(Y_{ig}-\alpha_\ell-\beta_\ell -\gamma_\ell \bar{D}^{(i)}_g\right)^2D_{ig}\right]
\end{align*}
which can be reparameterized as 
\begin{align*}
\E\left[\left(Y_{ig}-\alpha_0-\gamma_\ell \bar{D}^{(i)}_g\right)^2(1-D_{ig})\right]+\E\left[\left(Y_{ig}-\alpha_1 -\gamma_\ell \bar{D}^{(i)}_g\right)^2D_{ig}\right]
\end{align*}
where $\alpha_0=\alpha_\ell$ and $\alpha_1=\alpha_\ell+\beta_\ell$. The first-order condition for $\alpha_0$ and $\alpha_1$ are:
\begin{align*}
0&=\E\left[\left(Y_{ig}-\alpha_0-\gamma_\ell \bar{D}^{(i)}_g\right)(1-D_{ig})\right],\quad 0=\E\left[\left(Y_{ig}-\alpha_1-\gamma_\ell \bar{D}^{(i)}_g\right)D_{ig}\right] 
\end{align*}
from which $\alpha_0=\E[Y_{ig}|D_{ig}=0]-\gamma_\ell \E[\bar{D}^{(i)}_g|D_{ig}=0]$, $\alpha_1=\E[Y_{ig}|D_{ig}=1]-\gamma_\ell \E[\bar{D}^{(i)}_g|D_{ig}=1]$. But $\beta_\ell=\alpha_1-\alpha_0$ and thus:
\begin{align*}
\beta_\ell&=\E[Y_{ig}|D_{ig}=1]-\E[Y_{ig}|D_{ig}=0]-\frac{\gamma_\ell}{n_g}(\E[S_{ig}|D_{ig}=1]-\E[S_{ig}|D_{ig}=0]).
\end{align*}
The first-order condition for $\gamma_\ell$ is:
\begin{align*}
0&=\E\left[\left(Y_{ig}-\alpha_0-\gamma_\ell \bar{D}^{(i)}_g\right)\bar{D}^{(i)}_g(1-D_{ig})\right]+\E\left[\left(Y_{ig}-\alpha_1-\gamma_\ell \bar{D}^{(i)}_g\right)\bar{D}^{(i)}_gD_{ig}\right]\\
&=\cov(Y_{ig},\bar{D}^{(i)}_g|D_{ig}=0)\P[D_{ig}=0]+\cov(Y_{ig},\bar{D}^{(i)}_g|D_{ig}=1)\P[D_{ig}=1]\\
&-\gamma_\ell\left(\V[\bar{D}^{(i)}_g|D_{ig}=0]\P[D_{ig}=0]+\V[\bar{D}^{(i)}_g|D_{ig}=1]\P[D_{ig}=1]\right)
\end{align*}
from which:
\[\gamma_\ell=\frac{\cov(Y_{ig},\bar{D}^{(i)}_g|D_{ig}=0)\P[D_{ig}=0]+\cov(Y_{ig},\bar{D}^{(i)}_g|D_{ig}=1)\P[D_{ig}=1]}{\V[\bar{D}^{(i)}_g|D_{ig}=0]\P[D_{ig}=0]+\V[\bar{D}^{(i)}_g|D_{ig}=1]\P[D_{ig}=1]}.\]
Next, $\V[\bar{D}^{(i)}_g|D_{ig}=d]=\frac{1}{n_g^2}\V[S_{ig}|D_{ig}=d]$ and
\begin{align*}
\cov(Y_{ig},\bar{D}^{(i)}_g|D_{ig}=d)&=\frac{1}{n_g}\cov(Y_{ig},S_{ig}|D_{ig}=d)\\
&=\frac{1}{n_g}\sum_{s=0}^{n_g}\E[Y_{ig}|D_{ig}=d,S_{ig}=s]\cov(\1(S_{ig}=s),S_{ig}|D_{ig}=d).
\end{align*}
But
\begin{align*}
\cov(\1(S_{ig}=s),S_{ig}|D_{ig}=d)&=\E[\1(S_{ig}=s)S_{ig}|D_{ig}=d]-\E[\1(S_{ig}=s)|D_{ig}=d]\E[S_{ig}|D_{ig}=d]\\
&=(s-\E[S_{ig}|D_{ig}=d])\P[S_{ig}=s|D_{ig}=d].
\end{align*}
Therefore,
\begin{align*}
\gamma_\ell&=\frac{\sum_{s=0}^{n_g}\E[Y_{ig}|D_{ig}=0,S_{ig}=s]n_g\P[D_{ig}=0](s-\E[S_{ig}|D_{ig}=0])\P[S_{ig}=s|D_{ig}=0]}{\V[S_{ig}|D_{ig}=0]\P[D_{ig}=0]+\V[S_{ig}|D_{ig}=1]\P[D_{ig}=1]}\\
&+\frac{\sum_{s=0}^{n_g}\E[Y_{ig}|D_{ig}=1,S_{ig}=s]n_g\P[D_{ig}=1](s-\E[S_{ig}|D_{ig}=1])\P[S_{ig}=s|D_{ig}=1]}{\V[S_{ig}|D_{ig}=0]\P[D_{ig}=0]+\V[S_{ig}|D_{ig}=1]\P[D_{ig}=1]}\\
&=\sum_{s=0}^{n_g}\phi_0(s)\E[Y_{ig}|D_{ig}=0,S_{ig}=s]+\sum_{s=0}^{n_g}\phi_1(s)\E[Y_{ig}|D_{ig}=1,S_{ig}=s]
\end{align*}
where
\[\phi_d(s)=\frac{n_g\P[D_{ig}=d]\P[S_{ig}=s|D_{ig}=d]}{\V[S_{ig}|D_{ig}=0]\P[D_{ig}=0]+\V[S_{ig}|D_{ig}=1]\P[D_{ig}=1]}\cdot (s-\E[S_{ig}|D_{ig}=d]).\]
Also note that:
\begin{align*}
\sum_{s=0}^{n_g}\phi_d(s)&=n_g\P[D_{ig}=d]\sum_{s=0}^{n_g}(s-\E[S_{ig}|D_{ig}=d])\P[S_{ig}=s|D_{ig}=d]=0.
\end{align*}
This implies that:
\begin{align*}
\sum_{s=0}^{n_g}\phi_d(s)\E[Y_{ig}|D_{ig}=d,S_{ig}=s]&=\sum_{s=0}^{n_g}\phi_d(s)(\E[Y_{ig}|D_{ig}=d,S_{ig}=s]-\E[Y_{ig}|D_{ig}=d,S_{ig}=0])\\
&+\sum_{s=0}^{n_g}\phi_d(s)\E[Y_{ig}|D_{ig}=d,S_{ig}=0]\\
&=\sum_{s=1}^{n_g}\phi_d(s)(\E[Y_{ig}|D_{ig}=d,S_{ig}=s]-\E[Y_{ig}|D_{ig}=d,S_{ig}=0])
\end{align*}
which gives the required result. $\square$


\paragraph{Proof of Corollary \ref{coro:lim}}

When the true model satisfies exchangeability, potential outcomes have the form $Y_{ig}(d,s)$. From Theorem \ref{thm:lim},
\begin{align*}
\gamma_\ell&=\sum_{s=1}^{n_g}\phi_0(s)\E[Y_{ig}(0,s)-Y_{ig}(0,0)]+\sum_{s=1}^{n_g}\phi_1(s)\E[Y_{ig}(1,s)-Y_{ig}(1,0)]
\end{align*}
By linearity, $\E[Y_{ig}(d,s)-Y_{ig}(d,0)]=s\kappa_d$ and thus:
\begin{align*}
\gamma_\ell&=\kappa_0\sum_{s=1}^{n_g}s\phi_0(s)+\kappa_1\sum_{s=1}^{n_g}s\phi_1(s).
\end{align*}
But
\begin{align*}
\sum_{s=1}^{n_g}s\phi_d(s)&=\frac{n_g\P[D_{ig}=d]\sum_{s=1}^{n_g}s(s-\E[S_{ig}|D_{ig}=d])\P[S_{ig}=s|D_{ig}=d]}{\V[S_{ig}|D_{ig}=0]\P[D_{ig}=0]+\V[S_{ig}|D_{ig}=1]\P[D_{ig}=1]}\\
&=\frac{n_g\P[D_{ig}=d]\V[S_{ig}|D_{ig}=d]}{\V[S_{ig}|D_{ig}=0]\P[D_{ig}=0]+\V[S_{ig}|D_{ig}=1]\P[D_{ig}=1]}.
\end{align*}
Hence,
\begin{align*}
\gamma_\ell&=\kappa_0\sum_{s=1}^{n_g}s\phi_0(s)+\kappa_1\sum_{s=1}^{n_g}s\phi_1(s)=n_g\kappa_1 \lambda + n_g\kappa_0(1-\lambda)
\end{align*}
where
\[\lambda=\frac{\P[D_{ig}=d]\V[S_{ig}|D_{ig}=d]}{\V[S_{ig}|D_{ig}=0]\P[D_{ig}=0]+\V[S_{ig}|D_{ig}=1]\P[D_{ig}=1]}.\]
But $n_g\kappa_d=\E[Y_{ig}(d,n_g)-Y_{ig}(d,0)]$ which gives the result for $\gamma_\ell$. On the other hand, from Theorem \ref{thm:dm}, the difference in means is:
\begin{align*}
\beta_\mathsf{D}&=\E[Y_{ig}|D_{ig}=1]-\E[Y_{ig}|D_{ig}=0]\\
&=\E[Y_{ig}(1,0)-Y_{ig}(0,0)]+\sum_{s=1}^{n_g}\E[Y_{ig}(1,s)-Y_{ig}(1,0)]\P[S_{ig}=s|D_{ig}=1]\\
&-\sum_{s=1}^{n_g}\E[Y_{ig}(0,s)-Y_{ig}(0,0)]\P[S_{ig}=s|D_{ig}=0].
\end{align*}
By linearity,
\begin{align*}
\beta_\mathsf{D}&=\E[Y_{ig}(1,0)-Y_{ig}(0,0)]+\kappa_1\sum_{s=1}^{n_g}s\P[S_{ig}=s|D_{ig}=1]-\kappa_0\sum_{s=1}^{n_g}s\P[S_{ig}=s|D_{ig}=0]\\
&=\E[Y_{ig}(1,0)-Y_{ig}(0,0)]+\kappa_1\E[S_{ig}|D_{ig}=1]-\kappa_0\E[S_{ig}|D_{ig}=0].
\end{align*}
But 
\begin{align*}
\beta_\ell&=\beta_\mathsf{D}-\frac{\gamma_\ell}{n_g}(\E[S_{ig}|D_{ig}=1]-\E[S_{ig}|D_{ig}=0])\\
&=\E[Y_{ig}(1,0)-Y_{ig}(0,0)]+(\kappa_1-\kappa_0)\{(1-\lambda)\E[S_{ig}|D_{ig}=1]+\lambda\E[S_{ig}|D_{ig}=0]\}
\end{align*}
which gives the required result. $\square$


\paragraph{Proof of Theorem \ref{thm:lim_inter}}

The coefficients from Equation \eqref{eq:lim_inter} are characterized by the minimization problem:
\begin{align*}
\min_{\tilde{\alpha}_\ell,\tilde{\beta}_\ell,\gamma^0_\ell,\gamma^1_\ell}\E\left[\left(Y_{ig}-\tilde{\alpha_\ell}-\tilde{\beta}_\ell D_{ig}-\gamma_\ell^0 \bar{D}^{(i)}_g(1-D_{ig})-\gamma_\ell^1 \bar{D}^{(i)}_gD_{ig}\right)^2\right]
\end{align*}
The objective function can be rewritten as:
\begin{align*}
\E\left[\left(Y_{ig}-\tilde{\alpha}_\ell-\gamma_\ell^0 \bar{D}^{(i)}_g\right)^2(1-D_{ig})\right]+\E\left[\left(Y_{ig}-\tilde{\alpha}_\ell-\tilde{\beta}_\ell-\gamma_\ell^1 \bar{D}^{(i)}_g\right)^2D_{ig}\right]
\end{align*}
which can be reparameterized as:
\begin{align*}
\E\left[\left(Y_{ig}-\alpha_0-\gamma_\ell^0 \bar{D}^{(i)}_g\right)^2(1-D_{ig})\right]+\E\left[\left(Y_{ig}-\alpha_1-\gamma_\ell^1 \bar{D}^{(i)}_g\right)^2D_{ig}\right]
\end{align*}
where $\alpha_0=\tilde{\alpha}_\ell$ and $\alpha_1=\tilde{\alpha}_\ell+\tilde{\beta}_\ell$. The first-order conditions for $\alpha_0$ and $\alpha_1$ imply: $\alpha_0=\E[Y_{ig}|D_{ig}=0]-\gamma^0_\ell \E[\bar{D}^{(i)}_g|D_{ig}=0]$ and $\alpha_1=\E[Y_{ig}|D_{ig}=1]-\gamma^1_\ell \E[\bar{D}^{(i)}_g|D_{ig}=1]$
and since $\tilde{\beta}_\ell=\alpha_1-\alpha_0$,
\begin{align*}
\tilde{\beta}_\ell&=\E[Y_{ig}|D_{ig}=1]-\E[Y_{ig}|D_{ig}=0]-\left(\frac{\gamma^1_\ell}{n_g}\E[S_{ig}|D_{ig}=1]-\frac{\gamma^0_\ell}{n_g}\E[S_{ig}|D_{ig}=0]\right).
\end{align*}
On the other hand, the first-order condition for each $\gamma^d_\ell$ implies:
\begin{align*}
0&=\E\left[\left(Y_{ig}-\E[Y_{ig}|D_{ig}=d]-\gamma^d_\ell (\bar{D}^{(i)}_g-\E[\bar{D}^{(i)}_g|D_{ig}=d])\right)\bar{D}^{(i)}_g\1(D_{ig}=d)\right]
\end{align*}
Thus:
\[\gamma^d_\ell=\frac{\cov(Y_{ig},\bar{D}^{(i)}_g|D_{ig}=d)}{\V[\bar{D}^{(i)}_g|D_{ig}=d]}\]
which gives the required result by calculations shown in the proof of Theorem \ref{thm:lim}. $\square$


\paragraph{Proof of Corollary \ref{coro:lim_inter}}

When the true model satisfies exchangeability, potential outcomes have the form $Y_{ig}(d,s)$. From Theorem \ref{thm:lim_inter},
\begin{align*}
\gamma^d_\ell&=\sum_{s=1}^{n_g}\omega_d(s)\E[Y_{ig}(d,s)-Y_{ig}(d,0)].
\end{align*}
By linearity, $\E[Y_{ig}(d,s)-Y_{ig}(d,0)]=s\kappa_d$ and thus:
\begin{align*}
\gamma^d_\ell&=\frac{\kappa_dn_g}{\V[S_{ig}|D_{ig}=d]}\sum_{s=1}^{n_g}s(s-\E[S_{ig}|D_{ig}=d])\P[S_{ig}|D_{ig}=d]=\kappa_d n_g\\
&=\E[Y_{ig}(d,n_g)-Y_{ig}(d,0)].
\end{align*}
On the other hand,
\begin{align*}
\tilde{\beta}_\ell&=\beta_\mathsf{D}-\left(\frac{\gamma^1_\ell}{n_g}\E[S_{ig}|D_{ig}=1]-\frac{\gamma^0_\ell}{n_g}\E[S_{ig}|D_{ig}=0]\right)\\
&=\E[Y_{ig}(1,0)-Y_{ig}(0,0)]+\kappa_1\E[S_{ig}|D_{ig}=1]-\kappa_0\E[S_{ig}|D_{ig}=0]\\
&-\kappa_1\E[S_{ig}|D_{ig}=1]+\kappa_0\E[S_{ig}|D_{ig}=0]\\
&=\E[Y_{ig}(1,0)-Y_{ig}(0,0)]
\end{align*}
as required. $\square$


\paragraph{Proof of Lemma \ref{lemma:sampsi}}
Take a constant $c\in\mathbb{R}$. Then
\begin{align*}
\P\left[\min_{\mathbf{a}\in\mathcal{A}_n}N(\mathbf{a})\le c\right]\le |\mathcal{A}_n|\max_{\mathbf{a}\in\mathcal{A}_n}\P[N(\mathbf{a})\le c].
\end{align*}
Now, for any $\delta>0$,
\begin{align*}
\P[N(\mathbf{a})\le c]&=\P\left[N(\mathbf{a})\le c,\left\vert\frac{\hat{\pi}(\mathbf{a})}{\pi(\mathbf{a})}-1\right\vert>\delta\right]+\P\left[N(\mathbf{a})\le c,\left\vert\frac{\hat{\pi}(\mathbf{a})}{\pi(\mathbf{a})}-1\right\vert\le \delta\right]\\
&\le \P\left[\left\vert\frac{\hat{\pi}(\mathbf{a})}{\pi(\mathbf{a})}-1\right\vert>\delta\right]+\P[N(\mathbf{a})\le c, G(n+1)\pi(\mathbf{a})(1-\delta)\le N(\mathbf{a})\le \pi(\mathbf{a})G(n+1)(1+\delta)]\\
&\le \P\left[\left\vert\frac{\hat{\pi}(\mathbf{a})}{\pi(\mathbf{a})}-1\right\vert>\delta\right]+\1(G(n+1)\pi(\mathbf{a})\le c/(1-\delta))\\
&\le \P\left[\left\vert\frac{\hat{\pi}(\mathbf{a})}{\pi(\mathbf{a})}-1\right\vert>\delta\right]+\1(G(n+1)\underline{\pi}_n\le c/(1-\delta))
\end{align*}
which implies
\begin{align*}
|\mathcal{A}_n|\max_{\mathbf{a}\in\mathcal{A}_n}\P[N(\mathbf{a})\le c]\le |\mathcal{A}_n|\max_{\mathbf{a}\in\mathcal{A}_n}\P\left[\left\vert\frac{\hat{\pi}(\mathbf{a})}{\pi(\mathbf{a})}-1\right\vert>\delta\right]+|\mathcal{A}_n| \1(G(n+1)\underline{\pi}_n\le c/(1-\delta))
\end{align*}
which converges to zero under condition \eqref{eq:cond_cells} using Lemma \ref{lemma:cons_pi}. $\square$


\paragraph{Proof of Theorem \ref{thm:inference}}

All the estimators below are only defined when $\1(N(\mathbf{a})>0)$. Because under the conditions for Lemma \ref{lemma:sampsi} this event occurs with probability approaching one, the indicator will be omitted to simplify the notation. Let $\varepsilon_{ig}(\mathbf{a})=Y_{ig}-\E[Y_{ig}|\mathbf{A}_{ig}=\mathbf{a}]$. 
I first show that $\max_{\mathbf{a}\in\mathcal{A}_n}|\hat{\mu}(\mathbf{a})-\mu(\mathbf{a})|=O_\P\left(\sqrt{\frac{\log|\mathcal{A}_n|}{G(n+1)\underline{\pi}_n}}\right)$. Letting $R_n=\sqrt{\frac{\log|\mathcal{A}_n|}{G(n+1)\underline{\pi}_n}}$, we need to show that for any $\eta>0$ there is a sufficiently large $M$ such that 
\[\P\left[\max_{\mathbf{a}\in\mathcal{A}_n}|\hat{\mu}(\mathbf{a})-\mu(\mathbf{a})|>M R_n\right]<\eta.\]
Since
\[\P\left[\max_{\mathbf{a}\in\mathcal{A}_n}|\hat{\mu}(\mathbf{a})-\mu(\mathbf{a})|>M R_n\right]=\E\left\{\P\left[\left.\max_{\mathbf{a}\in\mathcal{A}_n}|\hat{\mu}(\mathbf{a})-\mu(\mathbf{a})|>M R_n\right\vert \mathbf{A}\right]\right\}\]
it suffices to show that the probability on the right-hand side is less than $\eta$ for all $\mathbf{A}$.
Start by writing:
\begin{align*}
\hat{\mu}(\mathbf{a})-\mu(\mathbf{a})&=\frac{\sum_g\sum_i \varepsilon_{ig}(\mathbf{a})\1_{ig}(\mathbf{a})}{N(\mathbf{a})}\\
&=\frac{\sum_g\sum_i (\varepsilon_{ig}(\mathbf{a})\1(|\varepsilon_{ig}|>\xi_n)-\E[\varepsilon_{ig}(\mathbf{a})\1(|\varepsilon_{ig}|>\xi_n)|\mathbf{A}_{ig}])\1_{ig}(\mathbf{a})}{N(\mathbf{a})}\\
&\quad +\frac{\sum_g\sum_i (\varepsilon_{ig}(\mathbf{a})\1(|\varepsilon_{ig}|\le\xi_n)-\E[\varepsilon_{ig}(\mathbf{a})\1(|\varepsilon_{ig}|\le\xi_n)|\mathbf{A}_{ig}])\1_{ig}(\mathbf{a})}{N(\mathbf{a})}
\end{align*}
for some increasing sequence of constants $\xi_n$ whose rate will be determined along the proof. Let $\underline{\varepsilon}_{ig}(\mathbf{a})=\varepsilon_{ig}(\mathbf{a})\1(|\varepsilon_{ig}(\mathbf{a})|\le \xi_n)-\E[\varepsilon_{ig}(\mathbf{a})\1(|\varepsilon_{ig}(\mathbf{a})|\le\xi_n)|\mathbf{A}_{ig}]$ and $\bar{\varepsilon}_{ig}(\mathbf{a})=\varepsilon_{ig}(\mathbf{a})\1(|\varepsilon_{ig}(\mathbf{a})|>\xi_n)-\E[\varepsilon_{ig}(\mathbf{a})\1(|\varepsilon_{ig}(\mathbf{a})|>\xi_n)\mathbf{A}_{ig}]$. Then,
\begin{align*}
\P\left[\left.\max_{\mathbf{a}\in\mathcal{A}_n}|\hat{\mu}(\mathbf{a})-\mu(\mathbf{a})|>M R_n\right\vert \mathbf{A}\right]&=\P\left[\left.\max_{\mathbf{a}\in\mathcal{A}_n}\left\vert\frac{\sum_g\sum_i \underline{\varepsilon}_{ig}(\mathbf{a})\1_{ig}(\mathbf{a})}{N(\mathbf{a})}+\frac{\sum_g\sum_i \bar{\varepsilon}_{ig}(\mathbf{a})\1_{ig}(\mathbf{a})}{N(\mathbf{a})}\right\vert>M R_n\right\vert \mathbf{A}\right]\\
&\le |\mathcal{A}_n|\max_{\mathbf{a}\in\mathcal{A}_n}\P\left[\left.\frac{\sum_g\sum_i \underline{\varepsilon}_{ig}(\mathbf{a})\1_{ig}(\mathbf{a})}{N(\mathbf{a})}>\frac{M}{2}R_n\right\vert \mathbf{A}\right] \\
&+|\mathcal{A}_n|\max_{\mathbf{a}\in\mathcal{A}_n}\P\left[\left.\frac{\sum_g\sum_i \bar{\varepsilon}_{ig}(\mathbf{a})\1_{ig}(\mathbf{a})}{N(\mathbf{a})}>\frac{M}{2}R_n\right\vert \mathbf{A}\right].
\end{align*}
Now, by Markov's inequality,
\begin{align*}
\P\left[\left.\frac{\sum_g\sum_i \bar{\varepsilon}_{ig}(\mathbf{a})\1_{ig}(\mathbf{a})}{N(\mathbf{a})}>\frac{M}{2}R_n\right\vert \mathbf{A}\right]&\le \frac{1}{N(\mathbf{a})^2 \frac{M^2}{4}R_n^2}\sum_g\sum_i\E\left[|\varepsilon_{ig}(\mathbf{a})|^2\1(|\varepsilon_{ig}(\mathbf{a})|>\xi_n)|\mathbf{A}_{ig}=\mathbf{a}\right]\1_{ig}(\mathbf{a})\\
&= \frac{1}{N(\mathbf{a})^2 \frac{M^2}{4}R_n^2}\sum_g\sum_i\E\left[\frac{|\varepsilon_{ig}(\mathbf{a})|^{2+\delta}}{|\varepsilon_{ig}(\mathbf{a})|^{\delta}}\1(|\varepsilon_{ig}(\mathbf{a})|>\xi_n)|\mathbf{A}_{ig}=\mathbf{a}\right]\1_{ig}(\mathbf{a})\\
&\le \frac{4\bar{b}_2^{2+\delta}}{M^2}\cdot \frac{1}{\xi_n^\delta R_n^2 N(\mathbf{a})}.
\end{align*}
where $\bar{b}_2$ is a bound for the second moment of $\varepsilon_{ig}(\mathbf{a})$. Set $\xi_n=R_n^{-1}$. Then,
\begin{align*}
\frac{R_n^\delta}{R_n^2N(\mathbf{a})}&= \frac{R_n^\delta G(n+1)\underline{\pi}_n}{\log|\mathcal{A}_n|N(\mathbf{a})}=\frac{R_n^\delta}{\log|\mathcal{A}_n|}\cdot\frac{\underline{\pi}_n}{\pi(\mathbf{a})}\cdot \frac{\pi(\mathbf{a})}{\hat{\pi}(\mathbf{a})}\le \frac{R_n^\delta}{\log|\mathcal{A}_n|\min\limits_{\mathbf{a}\in\mathcal{A}_n}\left\{\frac{\hat{\pi}(\mathbf{a})}{\pi(\mathbf{a})}\right\}}.
\end{align*}
But
\begin{align*}
\left\vert \min_{\mathbf{a}\in\mathcal{A}_n}\left\{\frac{\hat{\pi}(\mathbf{a})}{\pi(\mathbf{a})}\right\}-1\right\vert=\left\vert \min_{\mathbf{a}\in\mathcal{A}_n}\left\{\frac{\hat{\pi}(\mathbf{a})}{\pi(\mathbf{a})}-1\right\}\right\vert\le \max_{{\mathbf{a}\in\mathcal{A}_n}}\left\vert \frac{\hat{\pi}(\mathbf{a})}{\pi(\mathbf{a})}-1\right\vert\to_\P 0
\end{align*}
by Lemma \ref{lemma:cons_pi2}. Thus,
\begin{align*}
\frac{R_n^\delta}{R_n^2N(\mathbf{a})}&\le \frac{R_n^\delta}{\log|\mathcal{A}_n|(1+o_\P(1))}.
\end{align*}
Setting $\delta=2$, we have that
\begin{align*}
|\mathcal{A}_n|\max_{\mathbf{a}\in\mathcal{A}_n}\P\left[\left.\frac{\sum_g\sum_i \bar{\varepsilon}_{ig}(\mathbf{a})\1_{ig}(\mathbf{a})}{N(\mathbf{a})}>\frac{M}{2}R_n\right\vert \mathbf{A}\right]&\le \frac{|\mathcal{A}_n|R_n^\delta}{\log|\mathcal{A}_n|(1+o_\P(1))}\\
&=\frac{4\bar{b}_2^{2+\delta}}{M^2}\cdot\frac{|\mathcal{A}_n|}{G(n+1)\underline{\pi}_n(1+o_\P(1))}
\end{align*}
which can be made arbitrarily small for sufficiently large $M$ given that $|\mathcal{A}_n|=O(G(n+1)\underline{\pi}_n)$. On the other hand, since $|\underline{\varepsilon}_{ig}(\mathbf{a})|\le 2\xi_n$, by Bernstein's inequality,

\begin{align*}
\P\left[\left.\frac{\sum_g\sum_i \underline{\varepsilon}_{ig}(\mathbf{a})\1_{ig}(\mathbf{a})}{N(\mathbf{a})}>\frac{M}{2}R_n\right\vert \mathbf{A}\right] &\le 2\exp\left\{-\frac{1}{8}\frac{M^2R_n^2 N(\mathbf{a})^2}{\sigma^2(\mathbf{a})N(\mathbf{a})+\xi_nMR_nN(\mathbf{a})/3}\right\}\\
&=2\exp\left\{-\frac{M^2}{8}\frac{R_n^2 N(\mathbf{a})}{\sigma^2(\mathbf{a})+M/3}\right\}\\
&\le 2\exp\left\{-\frac{M^2}{8}\frac{\log|\mathcal{A}_n|\min_{\mathbf{a}\in\mathcal{A}_n}\left\{\frac{\hat{\pi}_n(\mathbf{a})}{\pi(\mathbf{a})}\right\}}{\bar{\sigma}^2+M/3}\right\}\\
&=2\exp\left\{-\frac{M}{8}\frac{\log|\mathcal{A}_n|(1+o_\P(1))}{\frac{\bar{\sigma}^2}{M}+1/3}\right\}.
\end{align*}
Therefore,
\begin{align*}
|\mathcal{A}_n|\max_{\mathbf{a}\in\mathcal{A}_n}\P\left[\left.\frac{\sum_g\sum_i \underline{\varepsilon}_{ig}(\mathbf{a})\1_{ig}(\mathbf{a})}{N(\mathbf{a})}>\frac{M}{2}R_n\right\vert \mathbf{A}\right]&\le 2|\mathcal{A}_n|\exp\left\{-\frac{M}{8}\frac{\log|\mathcal{A}_n|(1+o_\P(1))}{\frac{\bar{\sigma}^2}{M}+1/3}\right\}\\
&=2\exp\left\{-\log|\mathcal{A}_n|\left(\frac{M(1+o_\P(1))}{\frac{8\bar{\sigma}^2}{M}+8/3}-1\right)\right\}
\end{align*}
which can be made arbitrarily small for sufficiently large $M$. This shows that $\max_{\mathbf{a}\in\mathcal{A}_n}|\hat{\mu}(\mathbf{a})-\mu(\mathbf{a})|=O_\P\left(\sqrt{\frac{\log|\mathcal{A}_n|}{G(n+1)\underline{\pi}_n}}\right)$. The proof to show that $\max_{\mathbf{a}\in\mathcal{A}_n}|\hat{\sigma}^2(\mathbf{a})-\sigma^2(\mathbf{a})|=O_\P\left(\sqrt{\frac{\log|\mathcal{A}_n|}{G(n+1)\underline{\pi}_n}}\right)$ follows the same argument, letting $u_{ig}(\mathbf{a})=\varepsilon^2_{ig}(\mathbf{a})-\sigma^2(\mathbf{a})$ and noting that:
\begin{align*}
\max_{\mathbf{a}\in\mathcal{A}_n}|\hat{\sigma}^2(\mathbf{a})-\sigma^2(\mathbf{a})|&=\max_{\mathbf{a}\in\mathcal{A}_n}\left\vert \frac{1}{N(\mathbf{a})}\sum_g\sum_i u_{ig}(\mathbf{a})\1_{ig}(\mathbf{a})\right\vert + \max_{\mathbf{a}\in\mathcal{A}_n}|\hat{\mu}(\mathbf{a})-\mu(\mathbf{a})|^2\\
&=\max_{\mathbf{a}\in\mathcal{A}_n}\left\vert \frac{1}{N(\mathbf{a})}\sum_g\sum_i u_{ig}(\mathbf{a})\1_{ig}(\mathbf{a})\right\vert+o_\P(R_n).
\end{align*}

Finally, for the last part,
\begin{align*}
\Delta&=\max_\mathbf{a\in\mathcal{A}_n}\sup_{x\in\mathbb{R}}\left\vert\P\left[\frac{\hat{\mu}(\mathbf{a})-\mu(\mathbf{a})}{\sqrt{\V[\hat{\mu}(\mathbf{a})|\mathbf{A}]}}\le x\right]-\Phi(x)\right\vert =\max_\mathbf{a\in\mathcal{A}_n}\sup_{x\in\mathbb{R}}\left\vert\E\left\{\P\left[\left.\frac{\hat{\mu}(\mathbf{a})-\mu(\mathbf{a})}{\sqrt{\V[\hat{\mu}(\mathbf{a})|\mathbf{A}]}}\le x\right\vert \mathbf{A}\right]-\Phi(x)\right\}\right\vert \\
&\le \E\left\{\max_\mathbf{a\in\mathcal{A}_n}\sup_{x\in\mathbb{R}}\left\vert\P\left[\left.\frac{\hat{\mu}(\mathbf{a})-\mu(\mathbf{a})}{\sqrt{\V[\hat{\mu}(\mathbf{a})|\mathbf{A}]}}\le x\right\vert \mathbf{A}\right]-\Phi(x)\right\vert\right\}
\end{align*}
and
\begin{align*}
\left\vert\P\left[\left.\frac{\hat{\mu}(\mathbf{a})-\mu(\mathbf{a})}{\sqrt{\V[\hat{\mu}(\mathbf{a})|\mathbf{A}]}}\le x\right\vert \mathbf{A}\right]-\Phi(x)\right\vert &=\left\vert\P\left[\left. \frac{\sum_g\sum_i \varepsilon_{ig}\1_{ig}(\mathbf{a})}{\sigma(\mathbf{a})\sqrt{N(\mathbf{a})}}\le x\right\vert \mathbf{A}\right]-\Phi(x)\right\vert.
\end{align*}
By the Berry-Esseen bound,
\begin{align*}
\sup_{x\in\mathbb{R}}\left\vert\P\left[\left. \frac{\sum_g\sum_i \varepsilon_{ig}\1_{ig}(\mathbf{a})}{\sigma(\mathbf{a})\sqrt{N(\mathbf{a})}}\le x\right\vert \mathbf{A}\right]-\Phi(x)\right\vert \le \frac{Cb^3}{\underline{\sigma}^3}\cdot \frac{1}{\sqrt{N(\mathbf{a})}}
\end{align*}
But $\frac{1}{N(\mathbf{a})}=O_\P\left(\frac{1}{G(n+1)\pi(\mathbf{a})}\right)$ and therefore,
\begin{align*}
\max_{\mathbf{a}\in\mathcal{A}_n}\sup_{x\in\mathbb{R}}\left\vert\P\left[\left. \frac{\sum_g\sum_i \varepsilon_{ig}\1_{ig}(\mathbf{a})}{\sigma(\mathbf{a})\sqrt{N(\mathbf{a})}}\le x\right\vert \mathbf{A}\right]-\Phi(x)\right\vert \le \frac{Cb^3}{\underline{\sigma}^3}\cdot O_\P\left(\frac{1}{\sqrt{G(n+1)\underline{\pi}_n}}\right)
\end{align*}
as required. $\square$


\paragraph{Proof of Theorem \ref{thm:bootstrap}}

We want to bound:
\[\Delta^*(\mathbf{a})=\sup_x \left\vert \P^*\left[\frac{\hat{\mu}^*(\mathbf{a})-\hat{\mu}(\mathbf{a})}{\sqrt{\V^*[\hat{\mu}(\mathbf{a})]}}\le x\right]-\Phi(x)\right\vert\]
uniformly over $\mathbf{a}$, where $\hat{\mu}^*(\mathbf{a})=\sum_g\sum_i Y^*_{ig}\1_{ig}(\mathbf{a})/N(\mathbf{a})$ if the denominator is non-zero, and zero otherwise, and where $Y^*_{ig}\1_{ig}(\mathbf{a})=(\bar{Y}(\mathbf{a})+(Y_{ig}-\bar{Y}(\mathbf{a}))w_{ig})\1_{ig}(\mathbf{a})=(\bar{Y}(\mathbf{a})+\hat{\varepsilon}_{ig}w_{ig})\1_{ig}(\mathbf{a})$. Then, if $N(\mathbf{a})>0$, $\E^*[\hat{\mu}^*(\mathbf{a})]=\hat{\mu}(\mathbf{a})$ and $\V^*[\hat{\mu}^*(\mathbf{a})]=\sum_g\sum_i \hat{\varepsilon}_{ig}^2\1_{ig}(\mathbf{a})/N(\mathbf{a})^2$. The centered and scaled statistic is given by:
\[\frac{\sum_g\sum_i \hat{\varepsilon}_{ig}\1_{ig}(\mathbf{a})w_{ig}}{\sqrt{\sum_g\sum_i \hat{\varepsilon}^2_{ig}\1_{ig}(\mathbf{a})}}.\]
By Berry-Esseen,
\begin{align*}
\sup_x \left\vert \P^*\left[\frac{\sum_g\sum_i \hat{\varepsilon}_{ig}\1_{ig}(\mathbf{a})w_{ig}}{\sqrt{\sum_g\sum_i \hat{\varepsilon}^2_{ig}\1_{ig}(\mathbf{a})}}\le x\right]-\Phi(x)\right\vert &\le C \frac{\sum_g\sum_i |\hat{\varepsilon}_{ig}|^3\1_{ig}(\mathbf{a})/N(\mathbf{a})}{\left(\sum_g\sum_i \hat{\varepsilon}^2_{ig}\1_{ig}(\mathbf{a})/N(\mathbf{a})\right)^{3/2}}\cdot \frac{1}{\sqrt{N(\mathbf{a})}}
\end{align*}
We also have that
\begin{align*}
\frac{\sum_g\sum_i |\hat{\varepsilon}_{ig}|^3\1_{ig}(\mathbf{a})}{N(\mathbf{a})} & \le \frac{\sum_g\sum_i |Y_{ig}-\mu(\mathbf{a})|^3\1_{ig}(\mathbf{a})}{N(\mathbf{a})}+|\bar{Y}(\mathbf{a})-\mu(\mathbf{a})|^3+O_\P(N(\mathbf{a})^{-2})\\
& = \E[|Y_{ig}-\mu(\mathbf{a})|^3]+O_\P(N(\mathbf{a})^{-1})
\end{align*}
and 
\begin{align*}
\frac{\sum_g\sum_i \hat{\varepsilon}_{ig}^2\1_{ig}(\mathbf{a})}{N(\mathbf{a})}&=\frac{\sum_g\sum_i (Y_{ig}-\mu(\mathbf{a}))^2\1_{ig}(\mathbf{a})}{N(\mathbf{a})}+(\bar{Y}(\mathbf{a})-\mu(\mathbf{a}))^2 \\
&=\sigma^2(\mathbf{a})+O_\P(N(\mathbf{a})^{-1})
\end{align*}
Then,
\begin{align*}
\Delta^*(\mathbf{a})&\le \sup_x \left\vert \P^*\left[\frac{\sum_g\sum_i \hat{\varepsilon}_{ig}\1_{ig}(\mathbf{a})w_{ig}}{\sqrt{\sum_g\sum_i \hat{\varepsilon}^2_{ig}\1_{ig}(\mathbf{a})}}\le x\right]-\Phi(x)\right\vert \1(N(\mathbf{a})>0)+2\1(N(\mathbf{a})=0)\\
&=C\frac{\E[|Y_{ig}-\mu(\mathbf{a})|^3]+O_\P(N(\mathbf{a})^{-1})}{\left[\sigma^2(\mathbf{a})+O_\P(N(\mathbf{a})^{-1}))\right]^{3/2}}\cdot \frac{\1(N(\mathbf{a})>0)}{\sqrt{N(\mathbf{a})}}+2\1(N(\mathbf{a})=0)
\end{align*}
and the result follows from the facts that $\P[\min_\mathbf{a}N(\mathbf{a})=0]\to 0$ and by Lemma \ref{lemma:cons_pi}. $\square$


\section{Proofs of Technical Lemmas}

\paragraph{Proof of Lemma \ref{lemma:cons_pi}} Take $\varepsilon>0$, then

\begin{align*}
|\mathcal{A}_n|\max_{\mathbf{a}\in\mathcal{A}_n}\P\left[\left\vert\frac{\hat{\pi}(\mathbf{a})}{\pi(\mathbf{a})}-1\right\vert>\varepsilon \right]&=|\mathcal{A}_n|\max_{\mathbf{a}\in\mathcal{A}_n}\P\left[\left\vert N(\mathbf{a})-\E[N(\mathbf{a})]\right\vert>\varepsilon\E[N(\mathbf{a})]  \right]
\end{align*}
Now, $N(\mathbf{a})-\E[N(\mathbf{a})]=\sum_g\sum_i \1_{ig}(\mathbf{a})-G(n+1)\pi(\mathbf{a})=\sum_g W_g$ where $W_g=\sum_i \1_{ig}(\mathbf{a})-(n+1)\pi(\mathbf{a})=N_g(\mathbf{a})-\E[N_g(\mathbf{a})]$. Note that the $W_g$ are independent, $\E[W_g]=0$ and:
\begin{align*}
|W_g|&\le (n+1)\max\{\pi(\mathbf{a}),1-\pi(\mathbf{a})\} \\
\V[W_g]&=\V\left[\sum_i \1_{ig}(\mathbf{a})\right]=\sum_i\V[\1_{ig}(\mathbf{a})]+2\sum_i\sum_{j>i}\cov(\1_{ig}(\mathbf{a}),\1_{jg}(\mathbf{a}))\\
&=(n+1)\pi(\mathbf{a})(1-\pi(\mathbf{a}))+(n+1)(n+2)\{\E[\1_{ig}(\mathbf{a})\1_{jg(\mathbf{a})}]-\pi(\mathbf{a})^2\}\\
&\le (n+1)\pi(\mathbf{a})(1-\pi(\mathbf{a}))+(n+1)(n+2)\pi(\mathbf{a})(1-\pi(\mathbf{a}))\\
&=(n+1)(n+3)\pi(\mathbf{a})(1-\pi(\mathbf{a})).
\end{align*}
Then, by Bernstein's inequality,
\begin{align*}
\P\left[\left\vert W_g]\right\vert>\varepsilon\E[N(\mathbf{a})]  \right]&\le 2\exp\left\{-\frac{\E[N(\mathbf{a})]^2\varepsilon^2}{\sum_g \V[W_g]+\frac{1}{3}(n+1)\max\{\pi(\mathbf{a}),1-\pi(\mathbf{a})\}\E[N(\mathbf{a})]\varepsilon}\right\} \\
&=2\exp\left\{-\frac{\frac{1}{2}G^2(n+1)^2\pi(\mathbf{a})^2\varepsilon^2}{G(n+1)(n+3)\pi(\mathbf{a})(1-\pi(\mathbf{a}))+\frac{1}{3}G(n+1)^2\pi(\mathbf{a})\max\{\pi(\mathbf{a}),1-\pi(\mathbf{a})\}\varepsilon}\right\}\\
&=2\exp\left\{-\frac{\frac{1}{2}G\pi(\mathbf{a})\varepsilon^2}{\frac{n+3}{n+1}(1-\pi(\mathbf{a}))+\frac{1}{3}\max\{\pi(\mathbf{a}),1-\pi(\mathbf{a})\varepsilon\}}\right\} \\
&\le 2\exp\left\{-\frac{\frac{1}{2}G\pi(\mathbf{a})\varepsilon^2}{\frac{n+3}{n+1}+\frac{\varepsilon}{3}}\right\}
\end{align*}
Therefore,
\begin{align*}
|\mathcal{A}_n|\max_{\mathbf{a}\in\mathcal{A}_n}\P\left[\left\vert\frac{\hat{\pi}(\mathbf{a})}{\pi(\mathbf{a})}-1\right\vert>\varepsilon \right]&\le 2\exp\left\{-G\underline{\pi}_n\left(\frac{\frac{1}{2}\varepsilon^2}{\frac{n+3}{n+1}+\frac{\varepsilon}{3}}-\frac{\log|\mathcal{A}_n|}{G\underline{\pi}_n}\right)\right\}\to 0.
\end{align*}
as required. $\square$

\paragraph{Proof of Lemma \ref{lemma:cons_pi2}} Take $\varepsilon>0$, then:
\begin{align*}
\P\left[\max_{\mathbf{a}\in\mathcal{A}_n}\left\vert\frac{\hat{\pi}(\mathbf{a})}{\pi(\mathbf{a})}-1\right\vert>\varepsilon \right]&\le \sum_{\mathbf{a}\in\mathcal{A}_n}\P\left[\left\vert\frac{\hat{\pi}(\mathbf{a})}{\pi(\mathbf{a})}-1\right\vert>\varepsilon \right] \le |\mathcal{A}_n|\max_{\mathbf{a}\in\mathcal{A}_n}\P\left[\left\vert\frac{\hat{\pi}(\mathbf{a})}{\pi(\mathbf{a})}-1\right\vert>\varepsilon \right]\to 0
\end{align*}
by Lemma \ref{lemma:cons_pi}. $\square$


\section{Proofs of Additional Results}\label{app:proofs}

\paragraph{Proof of Corollary \ref{coro:simplerand}}

Under exchangeability $\pi(\mathbf{a})=\pi(d,s)=p^d(1-p)^{1-d}{n \choose s}p^s(1-p)^{n-s}={n \choose s}p^{s+d}(1-p)^{n+1-s-d}$. This function is minimized at $\underline{\pi}_n=\underline{p}^{n+1}$ where $\underline{p}=\min\{p,1-p\}$. Thus,
\begin{align*}
\frac{\log|\mathcal{A}_n|}{G\underline{p}^{n+1}}&=\exp\left\{-\log G\left(1+\frac{n+1}{\log G}\log\underline{p}-\frac{\log \log |\mathcal{A}_n|}{\log G}\right)\right\}
\end{align*}
and since $|\mathcal{A}_n|=2(n+1)$, this term converges to zero when $(n+1)/\log G\to 0$. On the other hand,
\begin{align*}
\frac{|\mathcal{A}_n|}{G(n+1)\underline{\pi}_n}=\frac{2}{G\underline{p}^{n+1}}\le \frac{2\log|\mathcal{A}_n|}{G\underline{p}^{n+1}}\to 0
\end{align*}
under the same condition. $\square$


\paragraph{Proof of Corollary \ref{coro:2SR-FM}}

Under exchangeability, $\pi(\mathbf{a})=\pi(d,s)=q_{d+s}\left(\frac{s+1}{n+1}\right)^d$ $\times\left(1-\frac{s}{n+1}\right)^{1-d}$. Under the assignment mechanism in Section \ref{app:2sr}, $\underline{\pi}_n=q_0$ and $q_0\ge \frac{1}{2\left(n+3\right)}$ and thus:
\begin{align*}
\frac{\log|\mathcal{A}_n|}{G\underline{\pi}_n}&\le \frac{2(n+3)\log(2(n+2))}{G}= \exp\left\{-\log G\left(1-\frac{\log(2(n+2))}{\log G}-\frac{\log\log 2(n+1)}{\log G}\right)\right\}\to 0
\end{align*}
if $\log (n+1)/\log G\to 0$. Finally,
\[\frac{|\mathcal{A}_n|}{G(n+1)\underline{\pi}_n}\le \frac{4(n+1)(n+3)}{G}=\exp\left\{-\log G\left(1-\frac{\log(n+1)}{\log G}-\frac{\log(4(n+3))}{\log G}\right)\right\}\to 0\]
under the previous condition. $\square$

\end{appendices}

\end{document}

%% file: simul_results_300.txt
\begin{tabular}{lcccccc}
\hline\hline
\multicolumn{1}{l}{}&\multicolumn{1}{c}{$n = 2$}&\multicolumn{1}{c}{$n = 3$}&\multicolumn{1}{c}{$n = 4$}&\multicolumn{1}{c}{$n = 5$}&\multicolumn{1}{c}{$n = 6$}&\multicolumn{1}{c}{$n = 7$}\tabularnewline
\hline
{\bfseries Simple randomization}&&&&&&\tabularnewline
~~$(n+1)/\log(G)$&0.5260&0.7013&0.8766&1.0519&1.2273&1.4026\tabularnewline
~~Bias&-0.0022&-0.0003&0.0003&-0.0009&-0.0033&0.0011\tabularnewline
~~Variance&0.0027&0.0042&0.0070&0.0128&0.0214&0.0307\tabularnewline
~~95\% CI coverage - normal&0.9488&0.9460&0.9406&0.9344&0.9090&0.8872\tabularnewline
~~95\% CI length - normal&0.2041&0.2522&0.3238&0.4293&0.5516&0.6571\tabularnewline
~~95\% CI coverage - bootstrap&0.9504&0.9474&0.9430&0.9391&0.9221&0.9275\tabularnewline
~~95\% CI length - bootstrap&0.2044&0.2536&0.3291&0.4443&0.5891&0.7342\tabularnewline
~~Prop. empty cells&0.0000&0.0000&0.0000&0.0122&0.0946&0.3154\tabularnewline
~~$\E[N(0,0)]$&112&75&47&28&16&9\tabularnewline
~~$\E[N(0,n)]$&112&75&47&28&16&9\tabularnewline
\hline
{\bfseries Two-stage randomization}&&&&&&\tabularnewline
~~$\log(n+1)/\log(G)$&0.1926&0.2430&0.2822&0.3141&0.3412&0.3646\tabularnewline
~~Bias&0.0000&0.0003&0.0001&0.0000&-0.0005&0.0010\tabularnewline
~~Variance&0.0027&0.0031&0.0035&0.0039&0.0042&0.0046\tabularnewline
~~95\% CI coverage - normal&0.9518&0.9456&0.9472&0.9510&0.9468&0.9412\tabularnewline
~~95\% CI length - normal&0.2036&0.2166&0.2319&0.2422&0.2531&0.2616\tabularnewline
~~95\% CI coverage - bootstrap&0.9524&0.9470&0.9490&0.9512&0.9452&0.9424\tabularnewline
~~95\% CI length - bootstrap&0.2037&0.2172&0.2327&0.2432&0.2548&0.2636\tabularnewline
~~Prop. empty cells&0.0000&0.0000&0.0000&0.0000&0.0000&0.0002\tabularnewline
~~$\E[N(0,0)]$&113&100&89&82&76&72\tabularnewline
~~$\E[N(0,n)]$&112&100&88&82&76&72\tabularnewline
\hline
\end{tabular}

%% file: simul_results_600.txt
\begin{tabular}{lcccccc}
\hline\hline
\multicolumn{1}{l}{}&\multicolumn{1}{c}{$n = 2$}&\multicolumn{1}{c}{$n = 3$}&\multicolumn{1}{c}{$n = 4$}&\multicolumn{1}{c}{$n = 5$}&\multicolumn{1}{c}{$n = 6$}&\multicolumn{1}{c}{$n = 7$}\tabularnewline
\hline
{\bfseries Simple randomization}&&&&&&\tabularnewline
~~$(n+1)/\log(G)$&0.4690&0.6253&0.7816&0.9379&1.0943&1.2506\tabularnewline
~~Bias&-0.0004&0.0001&0.0002&0.0003&0.0010&0.0012\tabularnewline
~~Variance&0.0013&0.0020&0.0033&0.0058&0.0110&0.0186\tabularnewline
~~95\% CI coverage - normal&0.9522&0.9474&0.9514&0.9492&0.9381&0.9237\tabularnewline
~~95\% CI length - normal&0.1437&0.1766&0.2250&0.2958&0.3992&0.5171\tabularnewline
~~95\% CI coverage - bootstrap&0.9496&0.9484&0.9488&0.9506&0.9423&0.9388\tabularnewline
~~95\% CI length - bootstrap&0.1433&0.1767&0.2258&0.2989&0.4121&0.5456\tabularnewline
~~Prop. empty cells&0.0000&0.0000&0.0000&0.0000&0.0088&0.0958\tabularnewline
~~$\E[N(0,0)]$&225&150&94&56&33&19\tabularnewline
~~$\E[N(0,n)]$&225&150&94&56&33&19\tabularnewline
\hline
{\bfseries Two-stage randomization}&&&&&&\tabularnewline
~~$\log(n+1)/\log(G)$&0.1717&0.2167&0.2516&0.2801&0.3042&0.3251\tabularnewline
~~Bias&-0.0001&-0.0003&0.0009&-0.0003&0.0006&-0.0003\tabularnewline
~~Variance&0.0013&0.0015&0.0017&0.0019&0.0020&0.0022\tabularnewline
~~95\% CI coverage - normal&0.9506&0.9504&0.9502&0.9476&0.9442&0.9498\tabularnewline
~~95\% CI length - normal&0.1437&0.1526&0.1631&0.1695&0.1768&0.1817\tabularnewline
~~95\% CI coverage - bootstrap&0.9498&0.9488&0.9498&0.9460&0.9448&0.9490\tabularnewline
~~95\% CI length - bootstrap&0.1435&0.1523&0.1629&0.1694&0.1768&0.1818\tabularnewline
~~Prop. empty cells&0.0000&0.0000&0.0000&0.0000&0.0000&0.0000\tabularnewline
~~$\E[N(0,0)]$&225&200&177&164&152&144\tabularnewline
~~$\E[N(0,n)]$&225&200&177&164&152&144\tabularnewline
\hline
\end{tabular}

%% file: BO_hhsize.txt
\begin{tabular}{lr}
\hline\hline
\multicolumn{1}{l}{}&\multicolumn{1}{c}{Frequency}\tabularnewline
\hline
1&5,205\tabularnewline
2&1,410\tabularnewline
3&  168\tabularnewline
4&   15\tabularnewline
5&    1\tabularnewline
Total&6,799\tabularnewline
\hline
\end{tabular}

%% file: BO_ntreat.txt
\begin{tabular}{lr}
\hline\hline
\multicolumn{1}{l}{}&\multicolumn{1}{c}{Frequency}\tabularnewline
\hline
0&2,355\tabularnewline
1&3,782\tabularnewline
2&  607\tabularnewline
3&   52\tabularnewline
4&    3\tabularnewline
Total&6,799\tabularnewline
\hline
\end{tabular}

%% file: BO_regs_comparison.txt
\begin{tabular}{lccccccccccccc}
\hline\hline
\multicolumn{1}{l}{\bfseries }&\multicolumn{2}{c}{\bfseries Diff. Means}&\multicolumn{1}{c}{\bfseries }&\multicolumn{4}{c}{\bfseries Linear-in-Means}&\multicolumn{1}{c}{\bfseries }&\multicolumn{2}{c}{\bfseries Full}&\multicolumn{1}{c}{\bfseries }&\multicolumn{2}{c}{\bfseries Pooled}\tabularnewline
\cline{2-3} \cline{5-8} \cline{10-11} \cline{13-14}
\multicolumn{1}{l}{}&\multicolumn{1}{c}{coef}&\multicolumn{1}{c}{s.e.}&\multicolumn{1}{c}{}&\multicolumn{1}{c}{coef}&\multicolumn{1}{c}{s.e.}&\multicolumn{1}{c}{coef}&\multicolumn{1}{c}{s.e.}&\multicolumn{1}{c}{}&\multicolumn{1}{c}{coef}&\multicolumn{1}{c}{s.e.}&\multicolumn{1}{c}{}&\multicolumn{1}{c}{coef}&\multicolumn{1}{c}{s.e.}\tabularnewline
\hline
$D_{ig}$&0.006&0.016&&0.007&0.016&0.102**&0.042&&0.164**&0.066&&0.165**&0.065\tabularnewline
$\bar{D}^{(i)}_g$&&&&0.027&0.034&&&&&&&&\tabularnewline
$\bar{D}^{(i)}_g (1-D_{ig})$&&&&&&0.169**&0.068&&&&&&\tabularnewline
$\bar{D}^{(i)}_g D_{ig}$&&&&&&-0.064&0.039&&&&&&\tabularnewline
$\mathbbm{1}(S_{ig}=1) (1-D_{ig})$&&&&&&&&&0.146**&0.066&&&\tabularnewline
$\mathbbm{1}(S_{ig}=2) (1-D_{ig})$&&&&&&&&&0.14**&0.056&&&\tabularnewline
$\mathbbm{1}(S_{ig}=1) D_{ig}$&&&&&&&&&-0.041*&0.023&&&\tabularnewline
$\mathbbm{1}(S_{ig}=2) D_{ig}$&&&&&&&&&-0.051**&0.025&&&\tabularnewline
$\mathbbm{1}(S_{ig}>0) (1-D_{ig})$&&&&&&&&&&&&0.144**&0.06\tabularnewline
$\mathbbm{1}(S_{ig}>0) D_{ig}$&&&&&&&&&&&&-0.045**&0.02\tabularnewline
Constant&0.822***&0.013&&0.811***&0.024&0.756***&0.037&&0.706***&0.057&&0.705***&0.057\tabularnewline
Observations&&504&&&504&&504&&&504&&&504\tabularnewline
\hline
\end{tabular}

%% file: BO_regs_nonexchange_age.txt
\begin{tabular}{lcc}
\hline\hline
\multicolumn{1}{l}{}&\multicolumn{1}{c}{coef}&\multicolumn{1}{c}{s.e.}\tabularnewline
\hline
$D_{ig}$&0.165**&0.066\tabularnewline
$(1-D_{ig})D_{i1g}(1-D_{i2g})$&0.134**&0.067\tabularnewline
$(1-D_{ig})(1-D_{i1g})D_{i2g})$&0.162**&0.07\tabularnewline
$(1-D_{ig})D_{i1g}D_{i2g}$&0.14**&0.056\tabularnewline
$D_{ig}D_{i1g}(1-D_{i2g})$&-0.039&0.027\tabularnewline
$D_{ig}(1-D_{i1g})D_{i2g}$&-0.043*&0.026\tabularnewline
$D_{ig}D_{i1g}D_{i2g}$&-0.051**&0.025\tabularnewline
Constant&0.706***&0.057\tabularnewline
Observations&&504\tabularnewline
Chi-squared test&0.397&\tabularnewline
p-value&0.673&\tabularnewline
\hline
\end{tabular}

%% file: BO_regs_nonexchange_gender.txt
\begin{tabular}{lcc}
\hline\hline
\multicolumn{1}{l}{}&\multicolumn{1}{c}{coef}&\multicolumn{1}{c}{s.e.}\tabularnewline
\hline
$D_{ig}$&0.124***&0.045\tabularnewline
$\mathbbm{1}(S^m_{ig}=1) (1-D_{ig})$&0.035&0.036\tabularnewline
$\mathbbm{1}(S^m_{ig}=2) (1-D_{ig})$&0.105**&0.045\tabularnewline
$\mathbbm{1}(S^m_{ig}=1) D_{ig}$&-0.013&0.024\tabularnewline
$\mathbbm{1}(S^m_{ig}=2) D_{ig}$&-0.032&0.025\tabularnewline
$\mathbbm{1}(S^f_{ig}=1) (1-D_{ig})$&0.097***&0.035\tabularnewline
$\mathbbm{1}(S^f_{ig}=2) (1-D_{ig})$&0.101**&0.042\tabularnewline
$\mathbbm{1}(S^f_{ig}=1) D_{ig}$&-0.067***&0.025\tabularnewline
$\mathbbm{1}(S^f_{ig}=2) D_{ig}$&0.000&0.026\tabularnewline
Constant&0.745***&0.04\tabularnewline
Observations&&504\tabularnewline
Chi-squared test&1.578&\tabularnewline
p-value&0.182&\tabularnewline
\hline
\end{tabular}